\documentclass[12pt]{article}
\usepackage{amssymb,amsmath,fullpage}

\usepackage{graphicx}
\usepackage{todo}
\usepackage{nicefrac}
\usepackage{subfigure}

\newcommand{\sign}{\text{sign}}
\newcommand{\non}{\nonumber}
\newcommand{\ra}{\rightarrow }

\newcommand{\bes} {\begin{eqnarray*}}
\newcommand{\ees} {\end{eqnarray*}}
\newcommand{\ds}{\displaystyle}
\newcommand{\be}{\begin{eqnarray}}
\newcommand{\ee}{\end{eqnarray}}

\newcommand{\dd} \partial
\newcommand{\vep}{\varepsilon}
\newcommand{\eps}{\epsilon}
\newcommand{\ie}{{\it i.e.}\ }
\newcommand{\eg}{{\it e.g.}\ }
\newcommand{\etal}{{\it et al.}\ }

\newcommand{\ppr}{p^{(D)}}
\newcommand{\prz}{P^{(D)}_0}
\newcommand{\phirz}{\phi^{(D)}_0}
\newcommand{\fprz}{{\cal F}^{(D)}_{p,0}}

\newcommand{\DD}{{\cal D}}
\newcommand{\GG}{{\cal G}}
\newcommand{\HH}{{\cal H}}

\newcommand{\nb}{\bar{n}}
\newcommand{\pb}{\bar{p}}
\newcommand{\nt}{\bar{n}}
\newcommand{\pt}{\bar{p}}

\newcommand{\jn}{j_n}
\newcommand{\jp}{j_p}
\newcommand{\fn}{{\cal F}_n}
\newcommand{\fp}{{\cal F}_p}

\newcommand{\ppo}{p^{(o)}}
\newcommand{\nno}{n^{(o)}}
\newcommand{\jpo}{j_p^{(o)}}
\newcommand{\jno}{j_n^{(o)}}
\newcommand{\Eoz}{E_0^{(o)}}

\newcommand{\poz}{P^{(o)}_0}

\newcommand{\phoz}{\phi^{(o)}_0}
\newcommand{\fpoz}{{\cal F}^{(o)}_{p,0}}

\newcommand{\ppoz}{p^{(o)}_0}

\newcommand{\nd}{n^{(d)}}

\newcommand{\pdz}{P^{(d)}_0}
\newcommand{\phidz}{\phi^{(d)}_0}
\newcommand{\fpdz}{{\cal F}^{(d)}_{p,0}}

\newcommand{\z}{\zeta}
\newcommand{\V}{{\cal V}}
\newcommand{\Vp}{{\cal V}_+}
\newcommand{\Vm}{{\cal V}_-}
\newcommand{\Q}{{\cal Q}}
\newcommand{\Qdim}{{\cal Q}^{(\mbox{dim})}}
\newcommand{\Qp}{{\cal Q}_+}
\newcommand{\Qm}{{\cal Q}_-}
\newcommand{\Qpz}{{\cal Q}_{+,0}}
\newcommand{\Qmz}{{\cal Q}_{-,0}}
\newcommand{\Zo}{{\cal Z}_1}
\newcommand{\Zt}{{\cal Z}_2}

\makeatletter
\g@addto@macro\@floatboxreset\centering
\makeatother
\usepackage{color}

\begin{document}

\title{Systematic derivation of a surface polarization model for planar perovskite solar cells}

\author{N.E. Courtier\thanks{Mathematical Sciences, University of Southampton, SO17 1BJ, UK}, J.M. Foster\thanks{Department of Mathematics, University of Portsmouth, PO1 3HF, UK}, S.E.J. O'Kane\thanks{Department of Physics, University of Bath, BA2 7AY, UK},  A.B. Walker$^{\ddagger}$ \& G. Richardson$^{*}$
}

\maketitle

\begin{abstract}
\noindent
Increasing evidence suggests that the presence of mobile ions in perovskite solar cells can cause a current-voltage curve hysteresis. Steady state and transient current-voltage characteristics of a planar metal halide CH$_3$NH$_3$PbI$_3$ perovskite solar cell are analysed with a drift-diffusion model that accounts for both charge transport and ion vacancy motion. The high ion vacancy density within the perovskite layer gives rise to narrow Debye layers (typical width $\sim$2nm), adjacent to the interfaces with the transport layers, over which large drops in the electric potential occur and in which significant charge is stored. Large disparities between (I) the width of the Debye layers and that of the perovskite layer ($\sim$600nm) and (II) the ion vacancy density and the charge carrier densities motivate an asymptotic approach to solving the model, while the stiffness of the equations renders standard solution methods unreliable. We derive a simplified \textit{surface polarisation} model in which the slow ion dynamic are replaced by interfacial (nonlinear) capacitances at the perovskite interfaces. Favourable comparison is made between the results of the asymptotic approach and numerical solutions for a realistic cell over a wide range of operating conditions of practical interest.
\end{abstract}

\section{\label{intro}Introduction} 

Since the first use of methylammonium lead tri-halide perovskite as a sensitizer in a dye-sensitized solar cell \cite{Kojima09}, and its subsequent incorporation into a novel thin film solar technology as a bulk solar absorber \cite{Lee12,Kim12}, the efficiency of perovskite solar cells (PSCs) has increased extremely rapidly from around 3\% to above 20\% \cite{Correa17}, a level that is comparable to the standard crystalline silicon devices. This increase, along with advances in the material properties and stability of PSCs, makes this area of photovoltaic research a very hot topic \cite{Niu15,Stranks15}.

\begin{figure}
\scalebox{0.6}{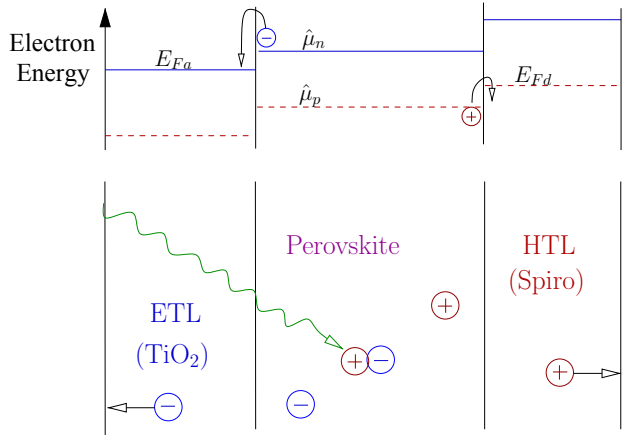}
\caption{(Top) Band diagram showing that holes preferentially move from the perovskite to the HTL and electrons to the ETL.
(Bottom) Schematic of a planar PSC showing photogeneration and transport of electrons and holes.}
\label{cellarchitecture}
\end{figure}

Typically PSCs contain a three-layer architecture consisting of a layer of semiconducting perovskite absorber sandwiched between a semiconducting hole-transport layer (HTL) and a semiconducting electron-transport layer (ETL), see figure \ref{cellarchitecture}. These transport layers are also referred to as selective or extraction layers or, alternatively, electron- and hole-blocking layers. A common pair of hole- and electron-transport materials are respectively Spiro-MeOTAD (2,2$^{\prime}$7,7$^{\prime}$-tetrakis-(N,N-di-p-methoxyphenyl amine)-9,9$^{\prime}$-spirobifluorene, here referred to as spiro) and titanium dioxide (TiO$_2$). Absorption of light occurs predominantly within the perovskite layer and is associated with the generation of an exciton which, due to its weak binding energy ($\sim50$eV)\cite{Koutselas96}, rapidly dissociates into a free electron in the conduction band, and a hole in the valence band, of the perovskite. These charge carriers move both in response to random thermal excitations (diffusion) and to internal electric fields (drift). The hole- and electron-transport materials are chosen such that their band energies give rise to a built-in electric field across the perovskite that separates the charge carriers. The field drives holes towards the HTL and electrons towards the ETL, generating a current at biases between zero and open circuit. Furthermore, the conduction band energy in the HTL is significantly above that in the perovskite, so that a potential barrier exists to the entry of electrons into this material from the perovskite. Similarly, the valence band energy in the ETL is significantly below that in the perovskite, so that a potential barrier exists to the entry of holes into this material from the perovskite. 

An unusual feature of PSCs is their long timescale transient behaviour occuring on the order of tens of seconds. This behaviour is exemplified by so-called current-voltage hysteresis \cite{Snaith14} whereby apparent hysteresis loops are observed in current-voltage (J-V) curves obtained by sweeping the voltage across a cell, from high to low and back again, and measuring the current as a function of voltage. From a practical point of view, this hysteresis has led to some unfortunate consequences, including inflated reports of power conversion efficiencies (PCEs) given that PCEs are often calculated from a current-voltage sweep. Long timescale transient behaviour has also been observed in dark current transients (whereby the cell is first held in the dark, then the applied voltage is suddenly changed and the resulting current measured)\cite{OKane17}. More recently, very long timescale transients lasting many hours have been observed in cell efficiency \cite{Domanski17}. These decays in PCE can be reversed by allowing the cell to recover in the dark. Various explanations have been proposed for these transient behaviours, including (a) large trap state densities close to the interfaces with the transport layers, (b) slow ferroelectric polarisation of the perovskite material and (c) the motion of iodide (I$^-$) vacancies within the perovskite material \cite{Snaith14}. As discussed in Richardson \etal\cite{Richardson16a}, it is now widely accepted that the only one of these mechanisms capable of explaining the data is iodide vacancy motion.

Various approaches may be used to model PSCs ranging from atomistic density functional theory (DFT) simulations, to drift-diffusion models of charge carrier and ion motion, to lumped parameter device models (equivalent circuits). DFT calculations, while perhaps the most fundamental approach, are so computationally intensive that they are incapable of describing the behaviour of a full cell. In practice they are used to obtain estimates of macroscopic quantities, such as ion vacancy densities and mobilities, from the atomistic structures of the materials forming the device\cite{Eames15}. In contrast, drift-diffusion models, which are applicable on the nanometre length scale and upward, describe the motion of electrons, holes and ion vacancies. Such models have been presented and solved in a number of works\cite{Richardson16a,Richardson16b,OKane17,Shen17,Calado16, Domanski17,Gottesman16,Vanreenen15,Neukom17}. However it is notable that, with the exception of two \cite{OKane17,Richardson16a}, all of these works use parameter
values that are very far from realistic. This may be ascribed to the extreme numerical stiffness of the problem owing to very narrow ($\approx$ 2 nm) Debye (boundary) layers that form as a result of ion accumulation/depletion at the edges of the perovskite layer.
In order to overcome this difficulty, Richardson \etal \cite{Richardson16a} adopted a combined numerical and asymptotic approach, in which the electrical properties of the Debye layers are modelled by a nonlinear surface (Debye layer) capacitance, based on estimates for the equilibrium ion vacancy density and mobility obtained from DFT calculations performed by Eames \etal\cite{Eames15}. The purpose of that work was to demonstrate that experimental J-V hysteresis data could be explained by the motion of ion vacancies in the perovskite layer and so the derivation of the asymptotic solution was not given there. 

The aim of this paper is to systematically derive the asymptotic approach used in the earlier work by Richardson \etal\cite{Richardson16a} and validate it against numerical solutions to the full model. A similar approach has been used for (i) asymptotic derivations of equivalent circuit models from drift-diffusion models \cite{Foster13,Foster14,Schmeiser94} (in the context of organic solar cells, PSCs and bipolar silicon devices, respectively); (ii) a matched asymptotic analysis of np-diodes \cite{Please82}; (iii) asymptotic derivations of the standard `regional' models of semiconductors from a drift-diffusion model \cite{Schmeiser93}; (iv) multidimensional models of bulk heterojunction solar cells \cite{Brinkman13,Richardson17}; and (v) the asymptotic analysis of quantum drift-diffusion models \cite{Black16}. Subsequent to Richardson \etal\cite{Richardson16a}, Ravishankar \etal \cite{Ravishankar17} published a heuristic model similar to the surface capacitance model used in this earlier work, which they term a {\it surface polarization model}. We argue that the systematic derivation of such models from the underlying drift-diffusion equations, as here, has the significant advantage of directly relating the surface capacitances to the device physics.

This work is set out as follows. In \S \ref{form}, we formulate the drift-diffusion model for a PSC, non-dimensionalise and estimate the model parameters. In \S \ref{asymp}, we use formal asymptotic methods, based on the parameter estimates made in \S \ref{form}, to derive a hierarchy of simplified models to the full PSC model including the surface polarization model of Ravishankar \etal\cite{Ravishankar17}. In \S \ref{compar}, the results of the simplified models are compared to numerical solutions of the full PSC model and finally, in \S \ref{concl}, we draw our conclusions.

\section{Problem formulation\label{form}}

Here we consider a perovskite absorber layer, sandwiched between an ETL and an HTL (typically TiO$_2$ and spiro, respectively). We make the assumption that the transport layers are sufficiently highly doped that they are effectively equipotential across their width and take the same potential as their respective contacts. In the perovskite, in line with DFT calculations on its chemical structure \cite{Eames15}, we assume there exists a high density of mobile anion vacancies, in addition to the charge carriers. The resulting dimensional model for the perovskite layer ($0<x<b$), following earlier work\cite{Richardson16a}, is outlined below.

\paragraph*{Dimensional model.} 
Conservation of holes (density $p$) and conduction electrons (density $n$) is described by
\be
\label{contperov}
\frac{\dd p}{\dd t} + \frac{1}{q} \frac{\dd \jp}{\dd x}  & = & G - R, \quad
\jp=-q {D}_p \left(\frac{\dd p}{\dd x} + \frac{p}{V_T} \frac{\dd \phi}{\dd x} \right), \nonumber  \\
\frac{\dd n}{\dd t}-    \frac{1}{q} \frac{\dd \jn}{\dd x} & = & G - R,   \quad
\jn=q {D}_n \left( \frac{\dd n}{\dd x} - \frac{n}{V_T} \frac{\dd \phi}{\dd x}  \right),
\ee
where $G$ is the photo-generation rate; $R(n,p)$ is the bulk recombination and thermal generation rate (henceforth abbreviated to recombination rate); $\phi$ is the electric potential; $\jn$ and $\jp$ are electron- and hole-currents, respectively; and $V_T=k T/q$ is the thermal voltage.
Similar equations for the conservation of positively-charged anion vacancies (density $P$) and negatively-charged cation vacancies (density $N$) take the form
\be
\label{chgeion}  
  \frac{\dd P}{\dd t} +  \frac{\dd \fp}{\dd x} &  =  & 0, \quad
 \fp=-{D}_+ \left(\frac{\dd P}{\dd x} + \frac{P}{V_T} \frac{\dd \phi}{\dd x} \right),  \nonumber  \\
 \frac{\dd N}{\dd t}+\frac{\dd \fn}{\dd x}  & = & 0,  \quad   \fn=-{D}_- \left( \frac{\dd N}{\dd x} - \frac{N}{V_T} \frac{\dd \phi}{\dd x}  \right),
\ee
where $\fp$ (and $\fn$) are the fluxes of the positive (and negative) ion vacancies (as opposed to the current fractions carried by these species). Both sets of equations couple to Poisson's equation for the electric potential
\be
\frac{\dd^2 \phi}{\dd x^2}= \frac{q}{\vep} \left( N-P +n-p  \right).
\ee
Boundary conditions at the edges of the perovskite, $x=0$ (the interface with the ETL) and $x=b$ (the interface with the HTL) take the form
\be
\label{dim-bcs}
\left. \begin{array}{c}
 n=n_0 \\*[3mm]
 \phi= \frac{V_{bi}-V_{ap}}{2}\\*[3mm]
 \jp= -q R_l \\
 \fn=0, \\
 \fp=0
\end{array} \right\}   x=0, \quad
\left. \begin{array}{c}
 p=p_0 \\*[3mm]
 \phi= -\frac{V_{bi}-V_{ap}}{2}\\*[3mm]
 \jn= -q R_r \\
 \fn=0, \\
 \fp=0
\end{array} \right\} x=b.   
\ee
where  $V_{ap}$ is the applied voltage; $V_{bi}$ is the built-in potential;  $R_l$ and $R_r$ are the interfacial charge recombination rates on $x=0$ and $x=b$, respectively; and the carrier densities on the interfaces are given by the expressions (see \eg Nelson \cite{Nelson03})
\bes
p_0= g_v \exp \left(\frac{\hat{\mu}_p- E_{F_d}}{k T} \right), \quad n_0=g_c \exp \left( \frac{E_{F_a}- \hat{\mu}_n}{k T} \right).
\ees
Here, $g_c$ and $g_v$ are the effective density of states in the conduction and valence bands of the perovskite, respectively; $\hat{\mu}_n$ and $\hat{\mu}_p$  are the perovskite conduction and valence band energies, respectively. In addition, we model the highly doped ETL and HTL  as metals in which $E_{F_d}$, the HOMO energy level of the HTL (Spiro), and $E_{F_a}$, the conduction band energy of the ETL (TiO$_2$), play the roles of the Fermi levels in these materials. These equations are supplemented by initial conditions, which we choose as follows to ensure charge neutrality,
\be
p|_{t=0}=p_0, \;  n|_{t=0}=n_0, \; N|_{t=0}=N_0, \; P|_{t=0}=N_0.
\ee

\paragraph*{The built-in voltage.} This quantity can be found from (\ref{contperov}) with boundary conditions (\ref{dim-bcs}) by noting that, at equilibrium, the photo-generation rate, applied voltage and electron- and hole-currents are all zero ($G=0$, $V_{ap}=0$ and $\jp=\jn=0$). The equilibrium solutions for $n$ and $p$ have the form
\bes
p=A \exp\left( -\frac{\phi}{V_T} \right), \quad n= B\exp\left( \frac{\phi}{V_T} \right),
\ees
in which the constants $A$ and $B$ are determined by the boundary conditions such that
\bes
p=p_0 \exp\left( -\frac{\phi}{V_T}-\frac{V_{bi}}{2 V_T} \right), \;\; n=n_0 \exp\left( \frac{\phi}{V_T}-\frac{V_{bi}}{2 V_T} \right).
\ees
Furthermore, since the rate of thermal generation and recombination must be equal ($R=0$) at equilibrium (see \eg (\ref{SRHequation})), we require $np = n_i^2$. It follows that
\be
V_{bi}=V_T \log \left( \frac{n_0 p_0}{n_i^2} \right),
\ee
which, with parameter estimates in Table 1, turns out to be 1V $\approx$ 39$V_T$.

\paragraph*{Recombination and photo-generation.} At the radiation intensities associated with sunlight, the bulk recombination rate within the perovskite, $R$, is believed to be predominantly trap assisted (although at higher radiation intensities bimolecular recombination becomes significant) \cite{Stranks14}. It is therefore appropriate to model bulk recombination by the Shockley-Read-Hall rate equation (see \eg Nelson\cite{Nelson03} {\S4.5.5})
\begin{equation}
\label{SRHequation}
R= {\frac{np-n_i^2}{\tau_pn+\tau_np+k_3}},    
\end{equation}
where $\tau_n$ and $\tau_p$ are the pseudo-lifetimes of conduction electrons and holes, respectively, and $k_3$ is a constant related to the pseudo-lifetimes and trap state energy level (typically negligible to the other terms in the denominator of (\ref{SRHequation}) when the cell is under illumination).
Furthermore, Stranks \etal\cite{Stranks14} suggest that bulk recombination is hole dominated ($\tau_p \gg \tau_n$), an assumption which is in line with that made in Richardson \etal\cite{Richardson16a}. There is still no consensus on the relative importance of interfacial recombination (at the interfaces between the perovskite and the transport layers) in comparison to bulk recombination although we note that this may be sample dependent. For example, de Quilettes \etal \cite{deQuilettes15} note that recombination within the perovskite occurs predominantly at crystal boundaries, which implies the magnitude of bulk recombination is strongly dependent upon the perovskite structure.

The photo-generation rate, $G$, is assumed to follow the Beer-Lambert law of light absorption; with light entering the device through the ETL (TiO$_2$). This has the form
\begin{equation}
G=F_{ph}\alpha\exp \left( -\alpha x \right),\label{Beer-Lambert}
\end{equation}
where $F_{ph}$ is the incident photon flux and $\alpha$ is the light absorption coefficient of the perovskite. 


\subsection{Non-dimensionalisation}
Dimensionless variables (denoted by a star) are introduced by rescaling (i) space with the width of the perovskite layer; (ii) voltages with the thermal voltage; (iii) charge carrier densities with the typical photo-generated charge density, $\Pi_0$ (see (\ref{chard})); (iv) current densities with the typical photo-generated current density, $q F_{ph}$; and (v) ion densities with the typical ion density, $N_0$. The rescaling reads
\begin{align}
\label{nondim}
&x = b x^*, &&\phi = V_T \phi^*, &&V_{ap}  =  V_T\Phi^* \nonumber \\ 
&t = \tau_{ion} t^*, && p = \Pi_0 p^*, && n = \Pi_0 n^* , \nonumber \\
&\jp = q F_{ph} {\jp}^*, && \jn = q F_{ph}{\jn}^* ,&&P = N_0 P^*, \nonumber \\
& N = N_0 N^*, &&\fp = \frac{D_+ N_0}{b} \fp^*, &&\fn = \frac{D_+ N_0}{b} \fn^*,\nonumber \\
& G = \frac{F_{ph}}{b}G^*, && R = \frac{F_{ph}}{b} R^* , &&R_{l,r}=F_{ph} R_{l,r}^*.
\end{align}
Here, $L_d$ is the Debye length calculated on the basis of the ion vacancy density and $\tau_{ion}$ is the characteristic timescale for ion motion defined, respectively, by
\bes
L_d = \left( \frac{\vep V_T}{ q N_0} \right)^{1/2}, \quad  \tau_{ion}=\frac{L_D b}{D_+}.
\ees
Furthermore, we take $\Pi_0$ to be the characteristic charge carrier density required to remove the photo-generated charge in the absence of an electric field
\bes
\Pi_0=\frac{F_{ph} b}{\hat{D}},
\label{chard}
\ees 
where $\hat{D}$ is a typical carrier diffusivity.
The non-dimensionalisation gives rise to the following dimensionless quantities that characterise the system:
\begin{align}
\label{parameters}
&\nu = \frac{D_+ b}{\hat{D} L_d}, &&\kappa_p = \frac{D_p}{\hat{D}}, && \kappa_n = \frac{D_n}{\hat{D}}, &&\nt = \frac{n_0}{\Pi_0}, \quad \pt = \frac{p_0}{\Pi_0}, \nonumber \\
&\gamma = \frac{b^2}{\hat{D} \tau_p}, &&\Delta = \frac{D_-}{D_+}, &&N_i = \frac{n_i}{\Pi_0}, &&\lambda = \frac{L_d}{b}, \quad \delta = \frac{\Pi_0}{N_0}, \nonumber \\
&\Phi_{bi}=\frac{V_{bi}}{V_T}, &&\Upsilon = \alpha b, &&\eps = \frac{\tau_n}{\tau_p}, &&K_3 = \frac{k_3}{\Pi_0\tau_p}.
\end{align}

\paragraph*{The dimensionless problem.} The system of equations obtained by applying the rescaling (\ref{nondim}) to the variables in (\ref{contperov})-(\ref{Beer-Lambert}) is
\be
\label{nondimDD}
\nu \frac{\dd p^*}{\dd t^*} + \frac{\dd \jp^*}{\dd x^*}  & = &  G^* - R^*, \quad
 \jp^*=-\kappa_p \left(\frac{\dd p^*}{\dd x^*} + p^* \frac{\dd \phi^*}{\dd x^*} \right),  \nonumber \\
  \nu \frac{\dd n^*}{\dd t^*}-  \frac{\dd \jn^*}{\dd x^*}  &= & G^* - R^*,  \quad
 \jn^*=\kappa_n \left( \frac{\dd n^*}{\dd x^*} - n^* \frac{\dd \phi^*}{\dd x^*}  \right), \nonumber \\ 
  \frac{\dd P^*}{\dd t^*} +  \lambda \frac{\dd \fp^*}{\dd x^*}  & = & 0, \qquad
 \fp^*=- \left(\frac{\dd P^*}{\dd x^*} +P^* \frac{\dd \phi^*}{\dd x^*} \right),\nonumber \\
  \frac{\dd N^*}{\dd t^*}+\lambda \frac{\dd \fn^*}{\dd x^*}  & = & 0, \qquad  \fn^*=-\Delta \left( \frac{\dd N^*}{\dd x^*} - N^*\frac{\dd \phi^*}{\dd x^*}  \right), 
  \nonumber   \\
\frac{\dd^2 \phi^*}{\dd {x^*}^2} & = & \frac{1}{\lambda^2} \left[ N^*-P^* +\delta \left(n^*-p^*\right) \right],
\ee \be
\label{nondimbcs}
\left. \begin{array}{c}
 n^* = \nb \\*[2mm]
 \ds \phi^* = \frac{\Phi_{bi}-\Phi^*}{2}
 \\*[2mm]
 \jp^* = -q R^*_l \\*[2mm]
 \fn^* =0, \\*[2mm]
 \fp^* =0
\end{array} \right\}   x^*=0,
\left. \begin{array}{c}
 p^* = \pb\\*[2mm]
\ds  \phi^*= -\frac{\Phi_{bi}-\Phi^*}{2}
 \\*[2mm]
 \jn^*= -q R_r^* \\*[2mm]
 \fn^*=0, \\*[2mm]
 \fp^*=0
\end{array} \right\} x^*=1, ~~~~~~~ 
\ee \begin{align}
& p^* = \pb, &&n^* = \nb, && N^* = 1, && P^* = 1 && \text{at } t^* = 0.
\label{nondimics}
\end{align}
The dimensionless recombination and generation rates (for a device under constant illumination) are given by 
\begin{align}
\label{recombgennd}
&R^*(n^*,p^*) = \gamma \left( \frac{n^*p^*-N_i^2}{n^*+\eps p^* +K_3} \right), &&G^*=\Upsilon \exp(-\Upsilon x^*).
\end{align}
Henceforth, we drop the star superscript from the dimensionless variables.


\subsection{\label{realparams}Parameter estimates for real devices}
A list of parameter estimates obtained from the literature is supplied in Table \ref{parameters table}. Note that $F_{ph}$, $\tau_n$, $\tau_p$ and $D_+$ are in line with the range of values found in the literature but have been specifically chosen to give good agreement to the experimental J-V curves presented by Richardson \etal\cite{Richardson16a}. Based on this data, the dimensionless parameters, corresponding to a cell with perovskite width $b=600$nm, are
\begin{align}
\label{param-estim}
&\lambda = 2.4 \times 10^{-3}, &&\nu= 5.8 \times 10^{-10}, &&\delta= 2.1\times 10^{-7}, \nonumber \\
&\kappa_n = \kappa_p = 1, &&\Delta=0, &&\eps= 3.3 \times 10^{-3}, \nonumber \\
&\pt = 0.30, &&\nt=20, &&N_i=8.6 \times 10^{-9}, \nonumber \\
&\gamma = 2.4, &&K_3 = 8.6 \times 10^{-9}, &&\Upsilon =  3.7 \;.
\end{align}
While, for a cell with perovskite width $b=150$nm, they remain unchanged except that
\begin{align*}
&\lambda = 1.0 \times 10^{-2}, &&\nu= 1.4 \times 10^{-10}, &&\delta=5.2\times 10^{-8}, \\
&\pt=1.2, && \nt=82, &&N_i=3.4 \times 10^{-8}, \\
&\gamma = 0.15, &&K_3 = 3.5 \times 10^{-8}, &&\Upsilon =  0.92 \;.
\end{align*}
For the range of possible perovskite layer thicknesses considered it always holds that $\delta \ll \lambda \ll 1$ and this observation motivates the asymptotic solution to the model considered in the next section.

\begin{table}[ht]
\begin{tabular}{ccc}
\hline
\textbf{Sym.} & \textbf{Description}  & \textbf{Value, Source} \\
\hline
$T$ & Temperature & 298 K\\
$F_{ph}$ & Incident photon flux & $9.5\times10^{20}$ m\textsuperscript{-2}s\textsuperscript{-1}, \cite{Loper14,Richardson16a}\\
$\alpha$ & Absorption coeff. & $6.1\times10^6$ m\textsuperscript{-1}, \cite{Loper14}\\
$b$ & Width & $1.5-6\times10^{-7}$ m, \cite{Lee12,Pockett15}\\
$V_{bi}$ & Built-in voltage & 1 V \\
$\hat{\mu}_n$ & Conduction band level & -3.7 eV, \cite{Schulz14}\\
$\hat{g}_c$ & Conduction band DoS & $8.1\times10^{24}$ m\textsuperscript{-3}, \cite{Brivio14}\\
$\hat{g}_v$ & Valence band DoS & $5.8\times10^{24}$ m\textsuperscript{-3}, \cite{Brivio14}\\
$\hat{\mu}_p$ & Valence band  level & -5.4 eV, \cite{Schulz14}\\
$\hat{D}_n$ & Electron diffusion coeff. & $1.7\times10^{-4}$ m\textsuperscript{2}s\textsuperscript{-1}, \cite{Stoumpos13}\\
$\hat{D}_p$ & Hole diffusion coeff. & $1.7\times10^{-4}$ m\textsuperscript{2}s\textsuperscript{-1}, \cite{Stoumpos13}\\
$D_+$ & Vacancy diffusion coeff. & $2.4 \times 10^{-16}$ m\textsuperscript{2}s\textsuperscript{-1},\cite{Eames15,Richardson16a}\\
$N_0$ & Vacancy density & $1.6\times10^{19}$ cm\textsuperscript{-3}, \cite{Walshangchem15}\\
$\varepsilon_p$ & Permittivity & $24.1\varepsilon_0$, \cite{Brivio14}\\
$\tau_n$ & Electron pseudo-lifetime & $3\times10^{-12}$ s, \cite{Richardson16a}\\
$\tau_p$ & Hole pseudo-lifetime & $9\times10^{-10}$ s, \cite{Richardson16a}\\
$g_c$ & TiO$_2$ conduction band DoS & $8.1\times10^{24}$ m\textsuperscript{-3}, \cite{Brivio14} \\
$g_v$ & Spiro valence band DoS & $5.8\times10^{24}$ m\textsuperscript{-3}, \cite{Brivio14}\\
$E_{Fa}$ & TiO$_2$ Fermi level & -4.0 eV, \cite{Schulz14}\\
$E_{Fd}$ & Spiro Fermi level & -5.0 eV, \cite{Schulz14} \\
\hline
\end{tabular}
\caption{Parameters for the device described in \ref{realparams}, where $\varepsilon_0$ is the permittivity of free space. Here $\alpha$ is calculated from Loper \etal\cite{Loper14} based on light wavelength of 585nm (close to the peak absorption of the perovskite layer). Unless stated otherwise, the parameters are for the perovskite layer.}
\label{parameters table}
\end{table}


\section{Asymptotic simplification of the model  ($\delta \ll \lambda \ll 1$)  \label{asymp}}
Here we assume dimensionless parameter sizes consistent with (\ref{param-estim}) and in particular require that $\delta \ll \lambda \ll 1$. In this scenario, the problem for the anion vacancy density and potential ($P$ and $\phi$) decouples from that for the charge carrier densities ($n$ and $p$) so that a very good estimate of $\phi$ can be obtained by ignoring the contributions of $n$ and $p$ in (\ref{nondimDD}). We shall further assume that the cation vacancies are effectively immobile on the timescales of interest, reflected in the choice of $\Delta=D_-/D_+=0$. This assumption, coupled to equations (\ref{nondimDD}) and initial conditions (\ref{nondimics}), imply that the cation vacancy density remains constant with $N \equiv 1$.

\subsection{The ion problem\label{id}} 
\begin{figure}
\centering
\scalebox{0.55}{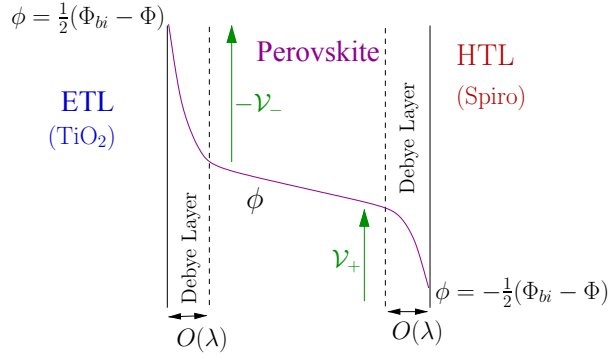}
\caption{Schematic representation of the Debye layers and the solution for the electric potential, $\phi$.} 
\label{deb_scheme}
\end{figure}

A good approximation to the potential can be obtained from the ion vacancy dependent equations in (\ref{nondimDD})-(\ref{nondimics}) at leading order, \ie
\be
\label{ionDDeq}
\frac{\dd P}{\dd t}+ \lambda \frac{\dd \fp}{\dd x} &= & 0, \quad \fp=- \left(\frac{\dd P}{\dd x} +P \frac{\dd \phi}{\dd x}   \right), 
\nonumber  \\
\frac{\dd^2 \phi}{\dd x^2} & =  &\frac{1}{\lambda^2} \left( 1-P \right),
\ee
with
\be
\label{ionDDbc}
\phi|_{x=0}  & = &  \frac{\Phi_{bi}-\Phi}{2}, \quad \left. \fp \right|_{x=0}=0,  \nonumber  \\
 \phi|_{x=1}  & =  & -\frac{\Phi_{bi}-\Phi}{2},  \quad \left. \fp \right|_{x=1}=0, \quad
P|_{t=0}=1.
\ee
Since $\lambda \ll 1$, these equations can be further approximated by using asymptotic boundary layer theory, in a similar vein to Richardson \etal\cite{Richardson09}. In the limit $\lambda \to 0$, the solution can be subdivided into three regions consisting of a bulk (or outer) region which is separated from the two boundaries by boundary layers of width $O(\lambda)$, see figure \ref{deb_scheme}. As is usual in this type of problem, these boundary layers are termed either Debye layers or double layers (we opt for the former usage).

\paragraph*{Bulk Region.} Away from the boundaries (\ie for $x \gg \lambda$ and $1-x \gg \lambda$) the variables 
$P$, $\fp$ and $\phi$ can be expanded, in powers of $\lambda$ and $\delta$, as follows:
\be
\label{ionvar}
P=1+ \cdots \;,\; \fp=\fpoz+ \cdots \;,\;   \phi=\phoz +\cdots.~~~~~~~ 
\ee
Substituting these expansions into (\ref{ionDDeq}), and assuming  $\delta/\lambda \ll 1$, gives, at leading order,
\bes
\frac{\dd \fpoz}{\dd x}=0, \qquad \fpoz=-\frac{\dd \phoz}{\dd x}.
\label{ionlead}
\ees
Note that correction terms in the expansions of $P$ and $\phi$ are $O(\delta)$ and $O(\delta/\lambda)$, respectively. These arise from the presence of the $O(\delta)$ charge carrier terms in Poisson's equation (last of (\ref{nondimDD})) and this is why the expansion breaks down if the value of either $n$ or $p$ becomes comparable to $O(\lambda/\delta)$.
It follows that  $\phi^{(o)}_{0,xx}=0$ and hence that
\be
\label{ionconcexp}
\phoz=W_-(t)\left(1-x\right) +W_+(t) x,
\ee
for arbitrary functions of time $W_-(t)$ and $W_+(t)$. It follows, on substituting into (\ref{ionlead}), that the leading order ion flux is given by
\be 
\label{ionfluxexp}
\fpoz=W_-(t)-W_+(t). 
\ee

\paragraph*{The Debye layers.} The asymptotic solution in the Debye layer about $x=0$ is obtained by rescaling space in the governing equations (\ref{ionDDeq})-(\ref{ionDDbc}) via 
\be
\label{zeta-def}
x=\lambda \zeta \;;
\ee
and substituting the asymptotic expansions 
\be
\label{deb-exp1}
P & = & \pdz(\z,t)+\cdots\;,\; \fp=\fpdz(\z,t)+\cdots \;,\; \nonumber \\
\phi & = & \phidz(\z,t)+\cdots
\ee
into the rescaled equations to obtain the leading order problem. The solution to which is given in Appendix \ref{append} and can be summarised as follows: (A) the leading order potential, $\phidz(\z,t)$, and vacancy distribution, $\pdz(\z,t)$, are both quasi-steady throughout the Debye layer, (B) the vacancy distribution is in quasi-equilibrium and so is Boltzmann distributed and (C) the potential satisfies a modified version of the Poisson-Boltzmann equation. The solution to this problem can be written in the form
\be
\label{debsoln1}
\pdz(\z,t)=\exp(-\theta(\z,\Vm(t))), \nonumber \\
\phidz(\z,t)=\theta(\z,\Vm(t))+W_-(t), 
\ee
where $W_-$ is the potential at the left-hand side of the bulk, to which $\phidz$ matches as $\z\ra +\infty$, and $\Vm(t)$ is the potential drop across the Debye layer (see figure \ref{deb_scheme}). The function $\theta(\z,\Vm)$ is defined  by the solution $\theta(z,\V)$ to the generic modified Poisson-Boltzmann problem
\be
 \label{thetaprob}
\frac{\dd^2 \theta}{\dd z^2}=1-e^{-\theta}, \quad  \theta|_{z=0}=-\V, \quad
\theta \ra 0,\, z \ra \infty.
\ee

Similarly, the asymptotic solution in the Debye layer about $x=1$ is obtained by rescaling space in the governing equations (\ref{ionDDeq})-(\ref{ionDDbc}) using the transformation
\be
x=1-\lambda \xi; \label{xi-def}
\ee
and substituting the asymptotic expansions 
\be
\label{deb-exp2}
P&=&\prz(\xi,t)+\cdots  \;,\;  \fp=\fprz(\xi,t)+\cdots \nonumber \\ 
\phi &=&\phirz(\xi,t)+\cdots  \;,\;  
\ee
into the resulting equations and solving at leading order. Once again this process is described in detail in Appendix \ref{append}. As in the other Debye layer, the leading order potential $\phirz(\xi,t)$ and vacancy distribution $\prz(\xi,t)$ can be written in the form
\be
 \label{debsoln2}
\prz(\z,t)=\exp(-\theta(\xi,\Vp(t)))  \nonumber \\ 
\phirz(\z,t)=\theta(\xi,\Vp(t))+W_+(t),
\ee
where $W_+$ is the potential at the right-hand side of the bulk (to which $\phirz$ matches as $\xi \ra +\infty$), $\Vp(t)$ is the potential drop across this Debye layer (see figure \ref{deb_scheme}) and $\theta(\xi,\Vp(t))$ is once again a solution to the problem (\ref{thetaprob}). 

In order to fully determine the leading order solutions in both Debye layers and the bulk region it is necessary to  solve for the time-dependent functions $\Vm$, $W_-$, $\Vp$ and $W_+$. The requirement that the leading order solutions in the Debye layers, (\ref{debsoln1}) and (\ref{debsoln2}), satisfy the potential boundary conditions in (\ref{ionDDbc}) gives
\be
 \label{debsoln3}
W_-(t)-\Vm(t)&=&\frac{\Phi_{bi}-\Phi(t)}{2}  \nonumber \\  
W_+(t)-\Vp(t)&=&-\frac{\Phi_{bi}-\Phi(t)}{2} \;. 
\ee

\begin{figure}
\includegraphics[width=0.6\columnwidth]{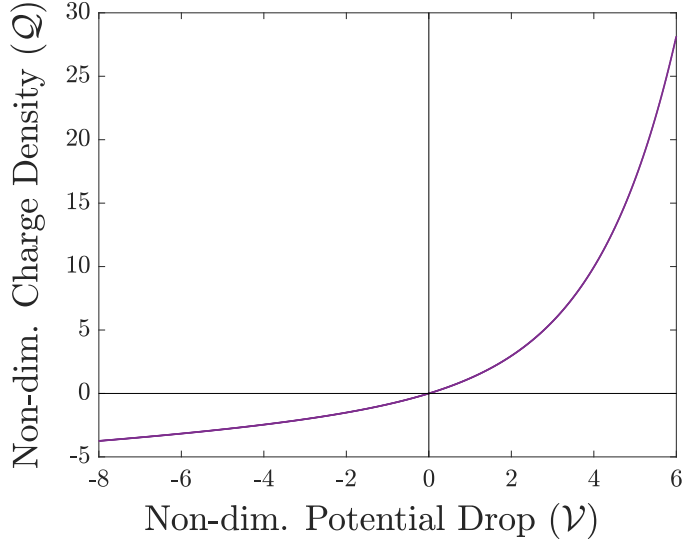}
\caption{Charge density in the Debye layer, ${\cal Q}$, vs. the potential drop across it, ${\cal V}$, defined by (\ref{qv-reln}), or equivalently (\ref{vq-reln}).}
\label{capac_plot}
\end{figure}

\paragraph*{Charge conservation within the Debye layers.}  A further two conditions on these four functions can be obtained by matching the flux of vacancies into the Debye layers with the leading order expansion of the vacancy conservation equations in the Debye layers. This leads to solvability conditions (described in Appendix \ref{append}) which can be interpreted in terms of global conservation of charge within the Debye layers. Since the leading order solutions for the vacancy densities within the Debye layers are quasi-steady, the total (dimensionless) charge per unit area within each Debye layer, ${\cal Q}$, can be related to the potential drop across the layer, $\V$, in the form of a nonlinear capacitance relation. Here the charges per unit area contained within each Debye layer ($\Qm$ in that about $x=0$ and $\Qp$ in that about $x=1$) are defined, in terms of the local Debye layer variables $\z$ and $\xi$, by
\be
\label{qm-def}
\Qm=\int_0^{\infty} \left(P-1\right) {\rm d}\z \;,\; \Qp=\int_0^{\infty} \left(P-1\right) {\rm d}\xi, 
\ee
and, as shown in Appendix \ref{append}, are related to the potential drops across the Debye layers ($\Vm$ and $\Vp$, respectively) via the capacitance relations
\be
\label{capacreln}
\Qm=Q(\Vm(t)) \;,\; \Qp=Q(\Vp(t)) \;,
\ee
where the function $Q(\V)$ is defined by
\be
\label{qv-reln}
Q(\V)= \sign(\V) \left( 2 \left( e^{\V}-1-\V \right)\right)^{1/2} \;.
\ee
This relation is plotted in figure \ref{capac_plot}. 
Furthermore, since vacancies (and hence charge) are conserved, the rate of change of the total charge per unit area within the Debye layers must equal the flux of (positively charged) vacancies flowing into each layer from the bulk region. Since the vacancy flux in the bulk region, $\fpoz$, is spatially independent, and given by (\ref{ionfluxexp}), this observation corresponds to the conditions
\bes
\frac{d \Qm}{d t}= W_+(t)-W_-(t) \;,\; \frac{d \Qp}{d t}= W_-(t)-W_+(t).
\ees
Alternatively, on eliminating $W_+$ and $W_-$ in favour of $\Vp$ and $\Vm$, we have the equivalent conditions
\be
\label{dQdt}
\frac{d \Qm}{d t} & = & -\left[\Phi_{bi}- \Phi(t)+\Vm(t)-\Vp(t)\right], \nonumber \\
\frac{d \Qp}{d t} & = & \Phi_{bi}- \Phi(t)+\Vm(t)-\Vp(t), 
\ee
which can be solved in conjunction with (\ref{capacreln})-(\ref{qv-reln}).
Adding these two equations together and integrating with respect to $t$ implies that the total charge in the Debye layers is conserved, \ie $\Qm(t)+\Qp(t)$ is constant. This is to be expected given that the predominant mobile charge carriers are the positive vacancies which cannot leave the perovskite region. Furthermore, since the net charge arising from both positive and negative vacancies will initially always be zero, it remains so for all time, \ie
\be
\label{zero-charge}
\Qm(t)=-\Qp(t).  
\ee
At this stage we can choose either to solve an ODE for $\Vp(t)$ or one for $\Qp(t)$. Since neither of these problems admit exact solution we opt to solve for $\Qp$ because this is preferable from a numerical point of view. We do this by noting that the inverse of (\ref{qv-reln}) is
\be
\label{vq-reln}
\V(Q)=\left\{ \begin{array}{cc}  
- \log_e \left( -\frac{1}{\mbox{LambertW}_0(-\exp(-( Q^2/2+1)))} \right) \nonumber \\
\mbox{for} \quad Q<0, \nonumber \\
 \log_e \left( -{\mbox{LambertW}_{-1}(-\exp(-( Q^2/2+1)))} \right) \\
 \mbox{for} \quad Q>0, \end{array} \right. 
\ee
where $\mbox{LambertW}_0(\cdot)$ and $\mbox{LambertW}_{-1}(\cdot)$ are the 0'th and $-1$'st  branch of the Lambert W function. On substituting  the above functional relation in (\ref{dQdt}), together with (\ref{zero-charge}), we obtain a single ODE for $\Qp(t)$:
\be
\label{qp-evol}
 \frac{{\rm d }\Qp}{{\rm d} t}= \Phi_{bi}- \Phi(t)+\V(-\Qp)-\V(\Qp)\;.
\ee
The solution to (\ref{qp-evol}) may be used to obtain the leading order bulk potential via (\ref{ionconcexp}), that is
\be
\phoz(x,t) = (&1-x)\left(\V(-\Qp(t)) +\frac{1}{2}\left[\Phi_{bi}-\Phi(t)\right] \right) \nonumber \\
&+ x \left( \V(\Qp(t))- \frac{1}{2}\left[\Phi_{bi}-\Phi(t)\right] \right).\label{phi-evol}
\ee

\paragraph*{Remark.} The dimensional surface charge density (in the Debye layers), $\Qdim$, is related to its non-dimensional counterpart, $\Q$, by
\bes
\Qdim = q \lambda b N_0 \Q.
\ees

\paragraph*{The uniformly valid approximation to $\phi$.} We can now write down a uniformly valid approximation to $\phi$ that is valid throughout the bulk and both Debye layers:
\be
\phi & \sim & (1-x)  \left(\V(-\Qp(t)) +\frac{1}{2}[\Phi_{bi}-\Phi(t)] \right) \nonumber \\
&& + x \left( \V(\Qp(t))- \frac{1}{2}[\Phi_{bi}-\Phi(t)] \right) \nonumber \\
&& + \theta\left(\frac{x}{\lambda}, \V(-\Qp(t)) \right) +\theta\left(\frac{1-x}{\lambda}, \V(\Qp(t))  \right),~~~~~~ \label{phi-approx}
\ee
where the function $\theta(z,\V)$ is defined implicitly in (\ref{theta-reln}) in Appendix \ref{append}. The corresponding uniformly valid asymptotic approximation for the anion vacancy density, $P$, is
\be
P \sim \exp \left( -\theta\left(\frac{x}{\lambda}, \V(-\Qp(t)) \right) \right)~~~~~~~~~~~~~~ \non \\
\qquad +\exp \left( -\theta\left(\frac{1-x}{\lambda}, \V(\Qp(t))  \right) \right)-1\;.
\ee

\subsection{Asymptotic approximation to the charge carrier equations}
As we demonstrate in \S\ref{compar}, the potential is well-approximated by the solution to the ion problem
(\ref{ionDDeq})-(\ref{ionDDbc}) and is almost entirely unaffected by the carrier distributions. Furthermore, since the Debye layers are extremely thin, the effects of both photo-generation and recombination within these layers are negligible so that, from (\ref{nondimDD}), the electron and hole currents are to a good approximation spatially independent across these layers,
\be
j_p^{(d)} \approx j_p^{(d)}(t),\; j_n^{(d)} \approx j_n^{(d)}(t),\; j_p^{(D)} \approx j_p^{(D)}(t),\;
 j_n^{(D)} \approx j_n^{(D)}(t). \nonumber
\ee
Furthermore, in these narrow regions, electron and hole densities are in approximate quasi-thermal equilibrium. In particular, in the Debye layers close to $x$ = 0 and $x$ = 1 respectively
\bes
\frac{\dd \nd}{\dd \z} \sim \nd \frac{\dd \phidz}{\dd \z}, \qquad 
\frac{\dd \ppr}{\dd \xi} \sim -\ppr \frac{\dd \phirz}{\dd \xi} 
\ees
Referring to the boundary conditions (\ref{nondimbcs}), we find that
\bes
\nd \sim \nt \exp \left(\phidz-\frac{1}{2} (\Phi_{bi}-\Phi)  \right) \text{ near } x=0 ,\\ 
\ppr \sim \pt \exp \left(-\phirz-\frac{1}{2} (\Phi_{bi}-\Phi) \right) \text{ near } x=1 .
\ees
For the purposes of predicting the output current of the device, we need only determine the carrier concentrations within the bulk region. Matching conditions on the bulk carrier problems (for $n$ and $p$) are obtained from the far-field behaviour of the Debye layer solutions, namely
\bes
\nd \ra \nt \exp( \Vm(t)) \ \ \mbox{as} \ \  \z \ra +\infty,\\
\ppr \ra \pt \exp( -\Vp(t)) \ \ \mbox{as} \ \  \xi \ra +\infty.
\ees
The appropriate boundary conditions on the bulk carrier densities are thus
\be
\label{bcbulknp}
\left. \begin{array}{l}
\nno=\nt \exp( \Vm(t)) \\
\jpo=-R_l
\end{array}  \right\}   x=0^+ \nonumber \\
\left. \begin{array}{l}
\ppo = \pt \exp( -\Vp(t)) \\
\jno=-R_r
\end{array} \right\}  x=1^-.  
\ee
The corresponding equations for the carrier densities in the bulk, as obtained from (\ref{nondimDD}), and are, on taking the physically appropriate limit $\nu \ra 0$, 
\be
\label{bulknp}
 \frac{\dd \jpo}{\dd x}  & = & G - R(\nno,\ppo) \;,\; \nonumber \\
 \jpo & = & -\kappa_p \left(\frac{\dd \ppo}{\dd x} - \ppo \Eoz\right),   \nonumber \\
   \frac{\dd \jno}{\dd x} & = & -G + R(\nno,\ppo) \;,\;  \nonumber \\
 \jno & = & \kappa_n \left( \frac{\dd \nno}{\dd x} + \nno \Eoz \right), 
\ee
where $\Eoz(t)$ is the leading order bulk electric field defined by $\Eoz(t)=-\dd \phoz/\dd x$ and from (\ref{phi-evol}) is given by
\be
\Eoz(t)=\Vm(t)-\Vp(t)+\Phi_{bi}-\Phi(t). \label{eoz}
\ee
Hence the asymptotic approximation to the charge carrier problem can  be found from the solution of (\ref{bcbulknp})-(\ref{bulknp}) in which the electric field term, $\Eoz(t)$, depends, via (\ref{eoz}), on the solution $\Qp(t)$ to the ion problem, through the relations $\Vm=\V(-\Qp)$ and $\Vp=\V(\Qp)$ (where the function $\V(\cdot)$ is defined in (\ref{vq-reln})). Usually the solution will have to be obtained numerically because of the nonlinearity of the recombination term. Nonetheless, numerically solving this reduced problem is considerably less challenging than directly tackling (\ref{nondimDD})-(\ref{nondimics}) because it excludes the Debye layers, over which the solution varies very rapidly. Finally, we note that the net current density $j^{(o)}(t)=\jno(x,t)+\jpo(x,t)$ is independent of the spatial variable $x$ and so can be found simply by evaluating the sum of the electron and hole current densities at any point in the domain.


\subsection{\label{speccase}An analytic solution in the limit  $\eps \ra 0$ with zero interfacial recombination} 
It is notable that the parameter $\eps=\tau_n/\tau_p$ is typically small (we estimate, on the basis of earlier work\cite{Richardson16a}, $\eps \approx 3.3 \times 10^{-3}$) while the other parameters in the SRH recombination term (\ref {recombgennd}), $N_i$ and $K_3$, are both very small. These observations lead us to set $N_i \equiv 0$, $K_3 \equiv 0$ and to investigate the small $\eps$ limit. In which case, provided that $p/n$ is not large, $R(n,p)$ can be approximated by
\be
R(n,p) \sim \gamma p.
\ee
If we restrict our interest to the case where interfacial recombination is negligible (i.e. if we take $R_l \equiv 0$ and $R_r \equiv 0$), it follows that the equation for the hole density decouples from that for the electron density (see (\ref{bulknp})) and can be reformulated as the following linear equation for $\ppo$:
\bes
 \frac{\dd \jpo}{\dd x}  = \Upsilon \exp(-\Upsilon x) - \gamma \ppo \;,\;
 \frac{\dd \ppo}{\dd x} - \ppo \Eoz= -\frac{\jpo}{\kappa_p}.
\ees
These may be solved by eliminating $\jpo$ from the above  to obtain a second order constant coefficient linear inhomogeneous equation for $\ppoz$, namely
\bes
\frac{\dd^2 \ppo}{\dd x^2}-\Eoz \frac{\dd \ppo}{\dd x}- \frac{\gamma \ppo}{\kappa_p}= -\Upsilon \exp(-\Upsilon x).
\ees
This can be rewritten in the form
\be
\label{lop1}
\frac{\dd^2 \ppo}{\dd x^2}-(\beta_1(t)+\beta_2(t)) \frac{\dd \ppo}{\dd x}+\beta_1(t)\beta_2(t) \ppo
\nonumber \\
=-d \exp(-\Upsilon x), 
\ee
where 
\be
\beta_1(t)&=&\frac{\Eoz(t)}{2}+\frac{\left((\Eoz(t))^2+4 \gamma/\kappa_p \right)^{1/2}}{2},
\nonumber \\
\beta_2(t)&=&\frac{\Eoz(t)}{2}-\frac{\left((\Eoz(t))^2+4 \gamma/\kappa_p \right)^{1/2}}{2},
\nonumber \\
d&=& \frac{\Upsilon}{\kappa_p}.
\ee
On noting that $\Eoz(t)=\beta_1(t)+\beta_2(t)$, the boundary conditions (\ref{bcbulknp}) can be stated as
\be
\label{lop2}
\left. \frac{\dd \ppo}{\dd x} 
- \ppo ( \beta_1(t)+\beta_2(t) ) \right|_{x=0}=0 \nonumber \\
\ppo|_{x=1}=\pt \exp(-\Vp(t)). 
\ee
The solution to (\ref{lop1}) and (\ref{lop2}) is
\be
\label{pout}
\ppo(x,t)=
-\frac{d e^{-\Upsilon x}}
{(\Upsilon+\beta_1(t))(\Upsilon+\beta_2(t))} ~~~\nonumber \\
+{\cal A}(t) e^{\beta_1(t) x}+ {\cal B}(t) e^{\beta_2(t) x}, 
\ee
where
\be
{\cal A}(t)&=& \frac{\hat{\cal A}(t)}{{\cal D}(t)}  \;,\; \qquad {\cal B}(t)= \frac{\hat{\cal B}(t)}{{\cal D}(t)}\;,\nonumber \\
\hat{\cal A}(t) & = & \beta_1(t) \pt \exp(-\Vp(t)) \nonumber \\
&& -\frac{d \left( e^{\beta_2(t)} (\beta_1(t)+\beta_2(t) +\Upsilon ) -\beta_1(t) e^{-\Upsilon} \right) }
{(\Upsilon+\beta_1(t))(\Upsilon+\beta_2(t))} \;,\nonumber \\
\hat{\cal B}(t) &= &-\beta_2(t) \pt \exp(-\Vp(t))  \nonumber \\
&& -\frac{d \left( e^{\beta_1(t)} (\beta_1(t)+\beta_2(t) +\Upsilon ) -\beta_2(t) e^{-\Upsilon} \right) }{(\Upsilon+\beta_1(t))
 (\Upsilon+\beta_2(t))}\;, \nonumber \\
{\cal D}(t) &= &\beta_1(t)e^{\beta_1(t)}-\beta_2(t)e^{\beta_2(t)} .
\ee

\paragraph*{An expression for the total current in the device.}
An expression for the hole current density $\jpo$ is found by substituting the solution (\ref{pout}) for $\ppo$ into (\ref{bulknp}); this gives
\be
\jpo=-\kappa_p \left(d e^{-\Upsilon x}  
\frac{(\Upsilon+\beta_1(t)+\beta_2(t))}
{(\Upsilon+\beta_1(t))(\Upsilon+\beta_2(t))} \right. ~~~~ \nonumber \\
\left. -\beta_2(t) {\cal A}(t) e^{\beta_1(t) x}-\beta_1(t) {\cal B}(t) e^{\beta_2(t) x} \right) .\label{jpo}
\ee
The total current $J(t)=\jpo(x,t)+\jno(x,t)$ is determined from the condition that $\jno(1,t)=0$ which thus implies that 
$J(t)=\jpo(1,t)$. It follows that
\be
 \label{Jexpr}
J(t)=-\kappa_p \left( d e^{-\Upsilon } \frac{(\Upsilon+\beta_1(t)+\beta_2(t))}{(\Upsilon+\beta_1(t))(\Upsilon+\beta_2(t))} \right. \nonumber \\
\left. -\beta_2(t) {\cal A}(t) e^{\beta_1(t) }-\beta_1(t) {\cal B}(t) e^{\beta_2(t) } \right) .
\ee

\paragraph*{Asymptotic solution for the bulk electron density.} In order to monitor whether this asymptotic solution breaks down, it is useful to derive an asymptotic expression for the bulk electron density, $\nno$, while recalling that we require $\ppo/\nno \gg \eps$ in order for the validity of the expansion. The equations and boundary conditions for $\nno$ are, at leading order,
\be
\frac{\dd \nno}{\dd x} + (\beta_1(t)+\beta_2(t))  \nno  & = & 
\frac{1}{\kappa_n} (J(t)-\jpo)\;,\; \nonumber \\
\nno|_{x=0} &=&\nt \exp( \Vm(t)),
\ee
in which we once again write $\Eoz=\beta_1+\beta_2$ and where $\jpo$ is given by (\ref{jpo}). The solution to this problem is
\be
\nno &= & \nt e^{\Vm(t)-(\beta_1(t)+\beta_2(t))  x} 
 + \frac{J(t)}{\kappa_n \Eoz}\left(1- e^{-(\beta_1(t)+\beta_2(t))  x} \right)  \nonumber \\
&& +\DD(t) \left(e^{-\Upsilon x} - e^{-(\beta_1(t)+\beta_2(t))  x} \right) \nonumber \\
&& +\GG(t) \left(e^{\beta_1(t) x}- e^{-(\beta_1(t)+\beta_2(t))  x} \right)
\nonumber \\
&& +\HH(t) \left( e^{\beta_2(t) x}- e^{-(\beta_1(t)+\beta_2(t))  x} \right),
\ee
where time-dependent functions $\DD$, $\GG$ and $\HH$ are given by
\be
 \DD(t) &= & \frac{\kappa_p}{\kappa_n} \left( \frac{d (\Upsilon+\beta_1(t)+\beta_2(t))}{(\beta_1(t)+\beta_2(t)-\Upsilon)(\Upsilon+\beta_1(t))(\Upsilon+\beta_2(t))} \right),\nonumber \\
  \GG(t) &=&-\frac{\kappa_p \beta_2(t) {\cal A}(t)}{\kappa_n(2 \beta_1(t)+\beta_2(t))}\;,\; \nonumber \\
  \HH(t) &= &- \frac{\kappa_p \beta_1(t) {\cal B}(t)}{\kappa_n(\beta_1(t)+2\beta_2(t))}.
\ee


\section{Comparison between numerical and asymptotic solutions to the model \label{compar}}

In this section, we compare the results obtained from (i) a numerical solution to the full model, (\ref{nondimDD})-(\ref{nondimics}), to those obtained from (ii) a combined  asymptotic/numerical approach, in which the ion problem is solved asymptotically as in \S\ref{id}, and from (iii) the special case described in \S\ref{speccase} which is entirely based on asymptotic approximations. In particular, we show that the results from (ii) the combined asymptotic/numerical approach, adopted in an earlier work \cite{Richardson16a}, compare extremely favourably to (i) numerical solution of the model.

\subsection{Numerical methods}

In approach (i), we use the method of lines. A detailed description of the numerical scheme is given by Courtier \etal\cite{Courtier17}, here we restrict ourselves to a brief outline. The spatial derivatives in equations (\ref{nondimDD}) are treated using a finite difference approach that is second-order accurate in space, both on the internal and boundary points, and chosen in such a way that conservation of species is also exact up to second order. After application of the finite difference approximations, the problem is reduced to a system of differential algebraic equations (DAEs) in which the ODEs arise from the evolution equations for $P$, $n$, and $p$, in (\ref{nondimDD}), and the algebraic equations are a result of Poisson's equation for the potential. Solving systems of DAEs presents a challenging numerical problem which we tackle using the {\tt ode15s} routine in Matlab \cite{MATLAB}. Owing to rapid changes of the solution curves within the narrow Debye layers, we find that the problem is sufficiently stiff to require non-uniform grid spacing and the additional precision offered by Advanpix's Multiprecision Computing Toolbox \cite{mct17}.

In approach (ii), the system of equations requiring numerical treatment is that for the charge carriers in the bulk, (\ref{bcbulknp})-(\ref{bulknp}). Having taken the asymptotic limits $\delta$, $\lambda$ and $\nu \to 0$, the remaining problem is a second-order boundary value problem (BVP). Crucially, since asymptotic expressions have been derived for the narrow Debye layers, only the solution in the bulk needs to be resolved numerically. This problem exhibits significantly reduced stiffness and, as a result, a straightforward application of the {\tt bvp4c} routine in Matlab \cite{MATLAB} suffices.

\subsection{Results}
\begin{figure*}
\includegraphics[width=0.49\textwidth]{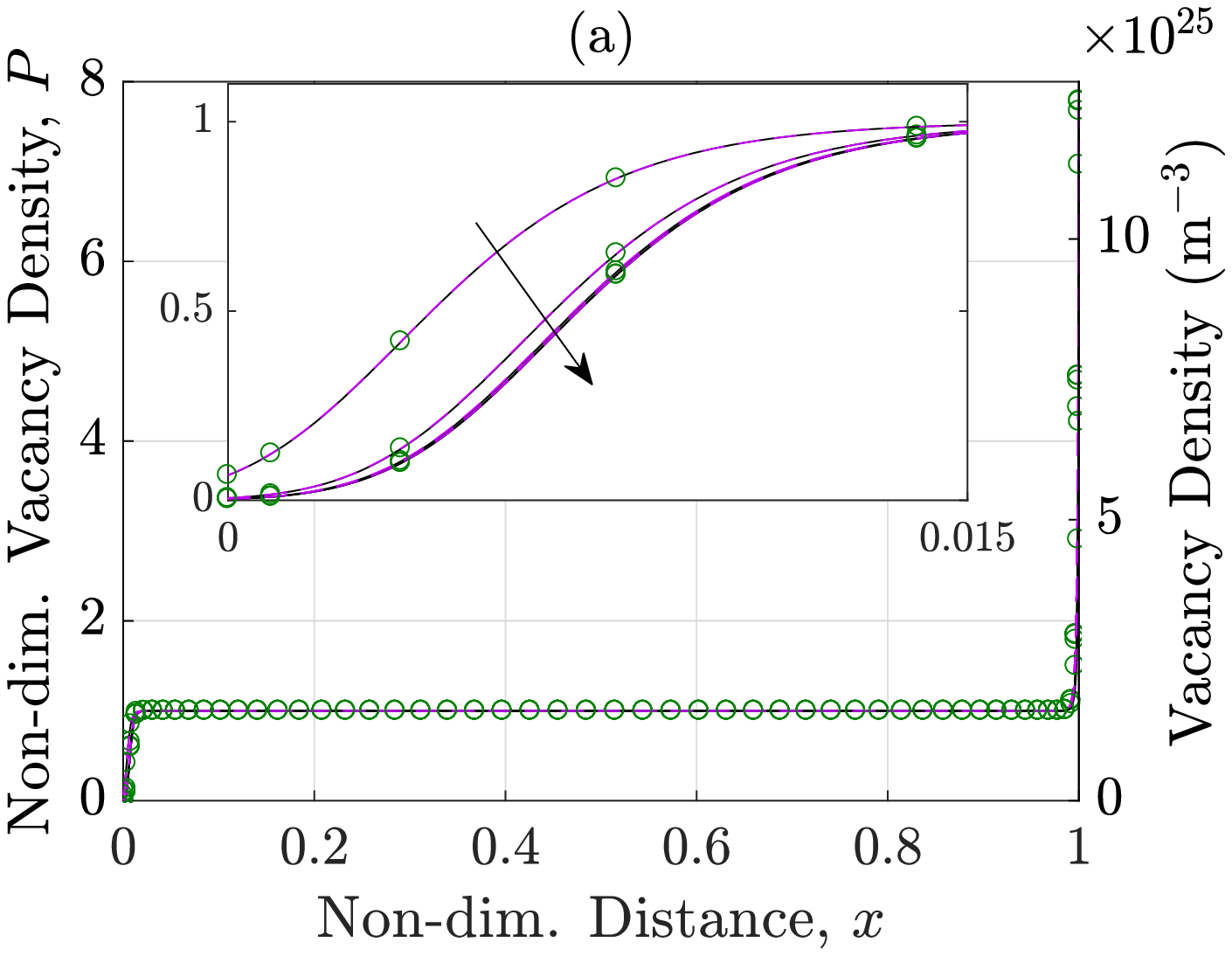}
\includegraphics[width=0.49\textwidth]{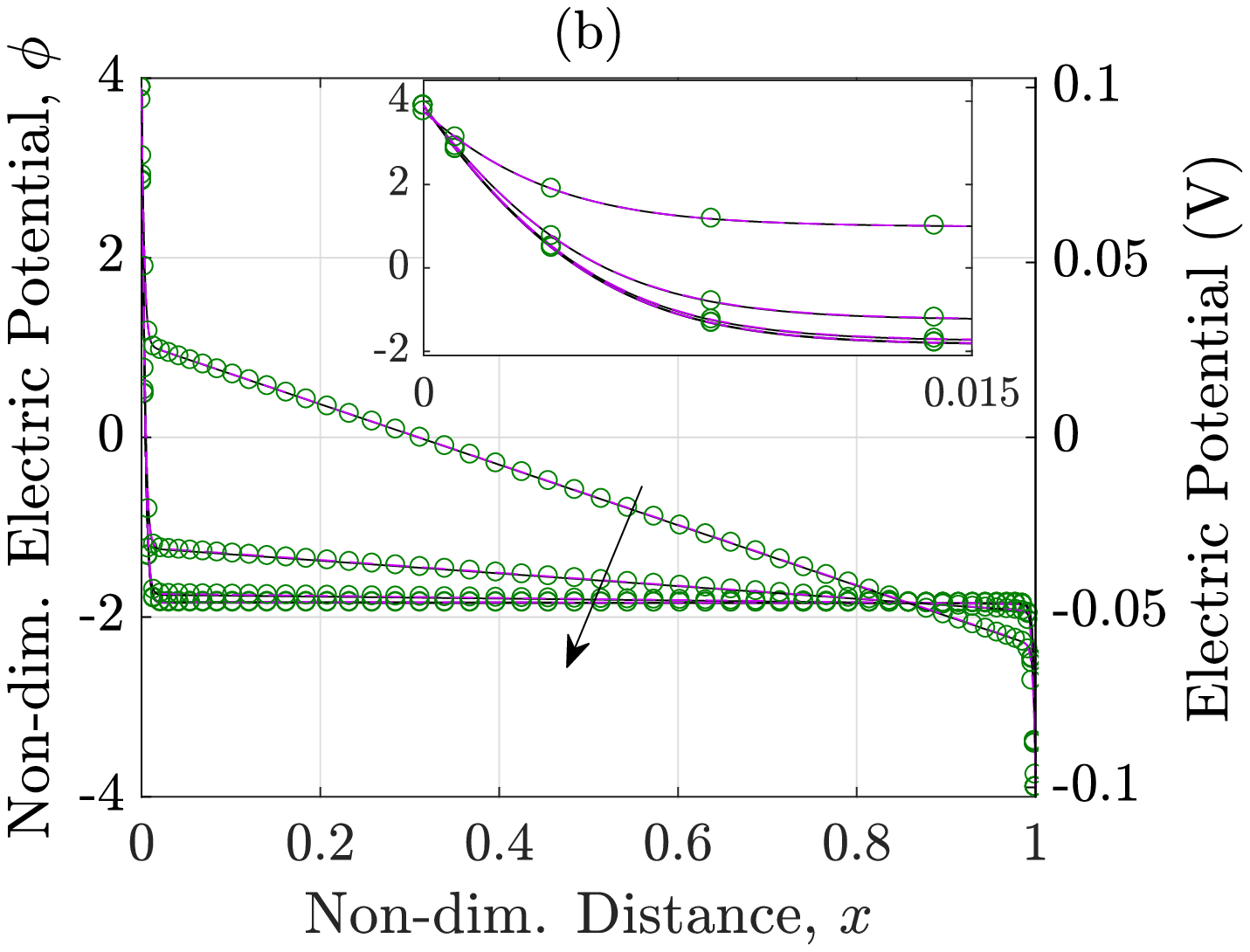} 
\includegraphics[width=0.49\textwidth]{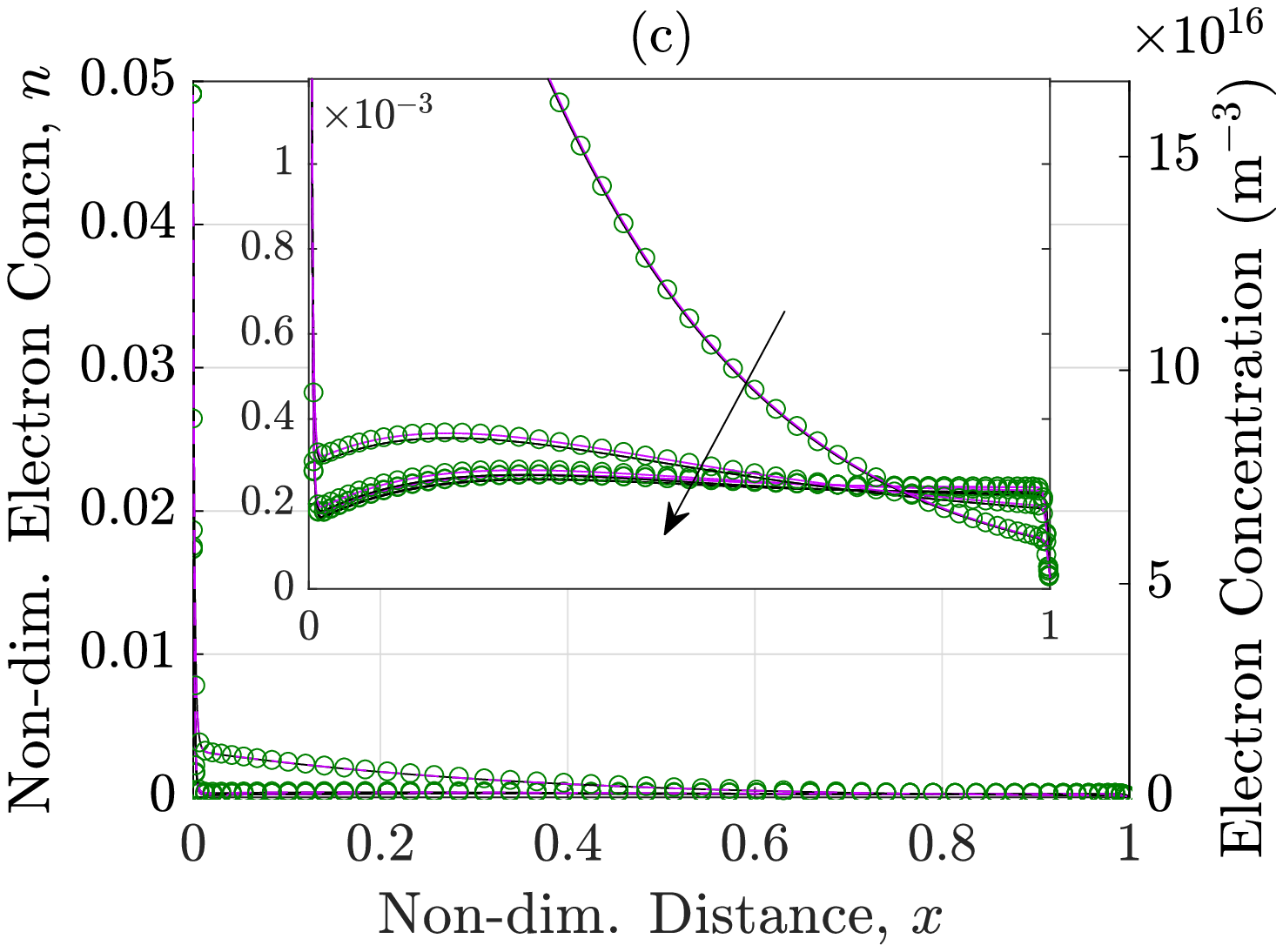} 
\includegraphics[width=0.49\textwidth]{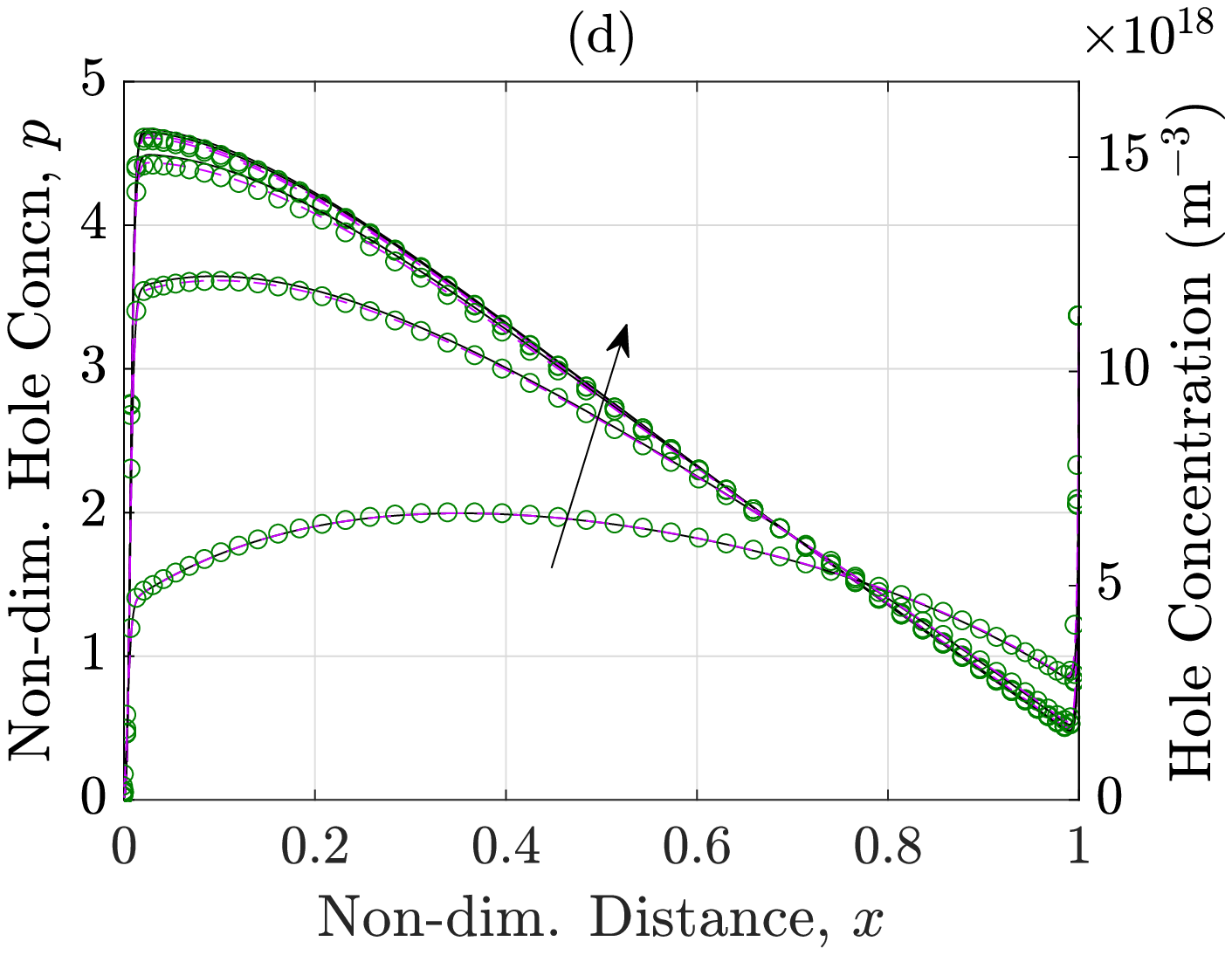} 
\caption{
(a) Anion vacancy density, (b) electric potential, (c) electron concentration and (d) hole concentration profiles across the perovskite layer of a cell during a smooth decrease of applied bias from $V_{ap}=V_{bi}$ to 0.8V. Insets focus on the left-hand (TiO$_2$/perovskite) boundary. Arrows indicate the direction of increasing time; black solid lines represent (i), the full numerical solutions, pink dashed lines represent (ii), the combined asymptotic/numerical approach and green circles represent (iii), the uniformly-valid asymptotic expansions from the fully asymptotic approach.}
\label{FastEvol-08V}
\end{figure*}

\begin{figure*}
\includegraphics[width=0.49\textwidth]{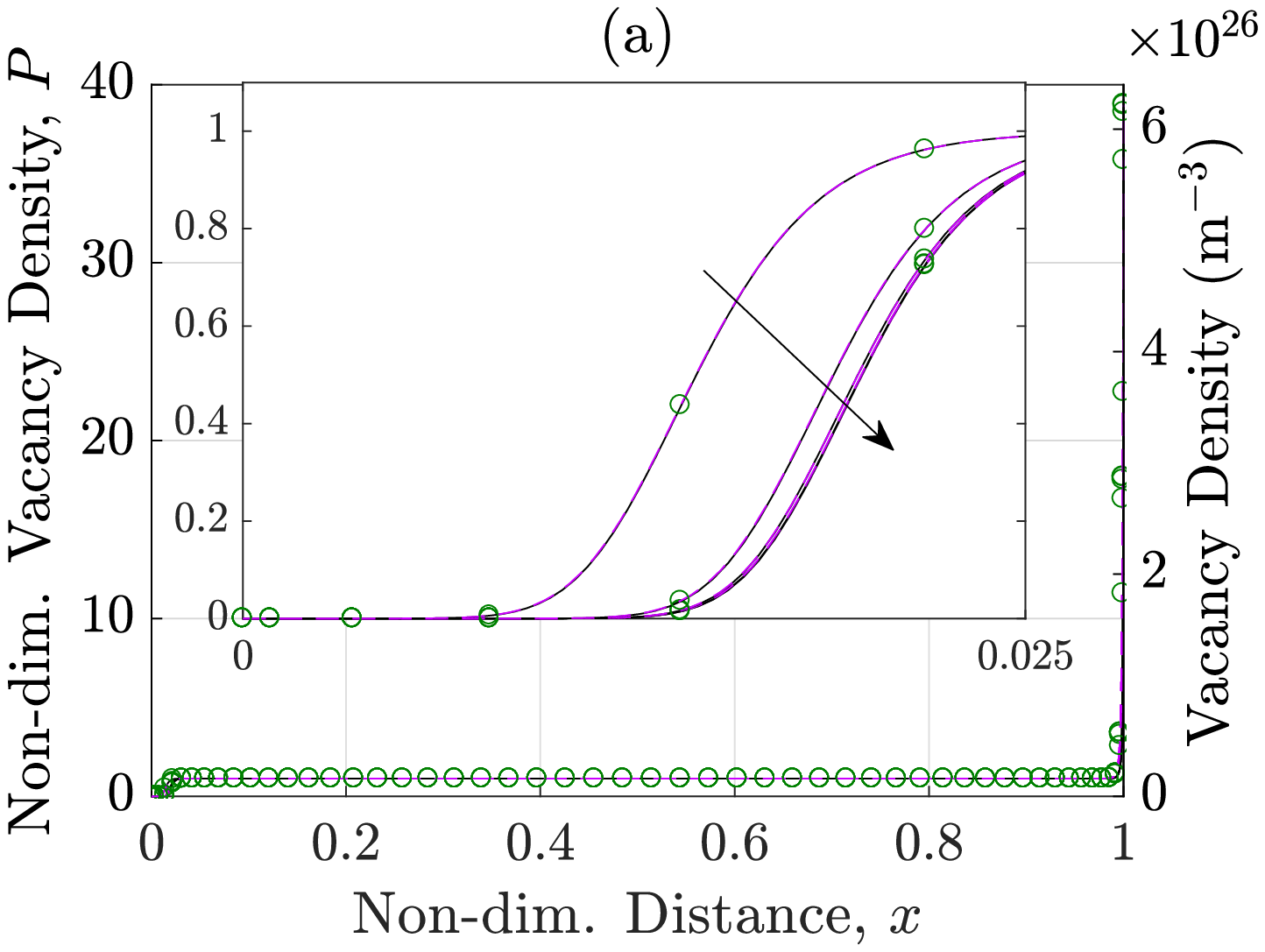}
\includegraphics[width=0.49\textwidth]{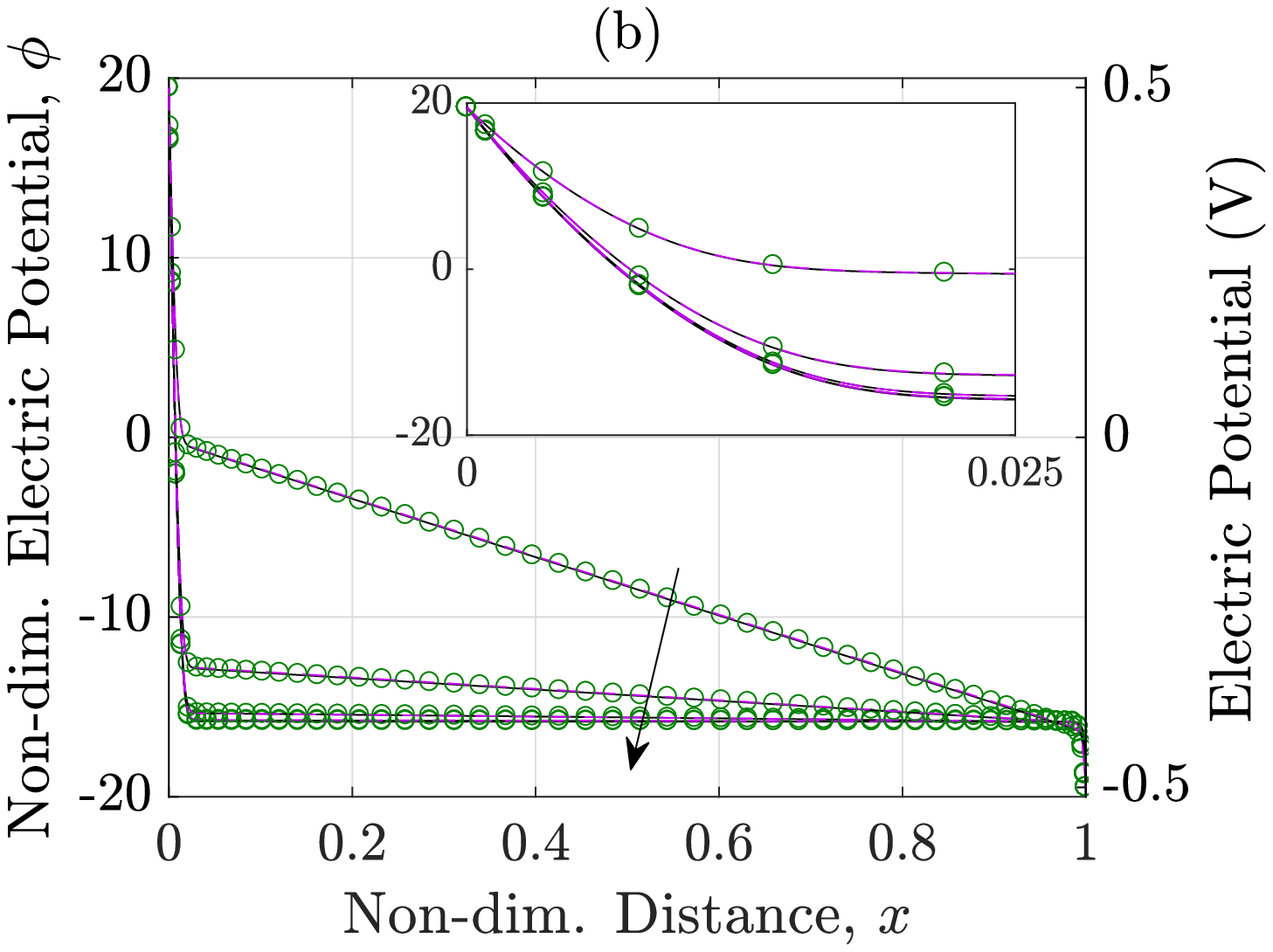}
\includegraphics[width=0.49\textwidth]{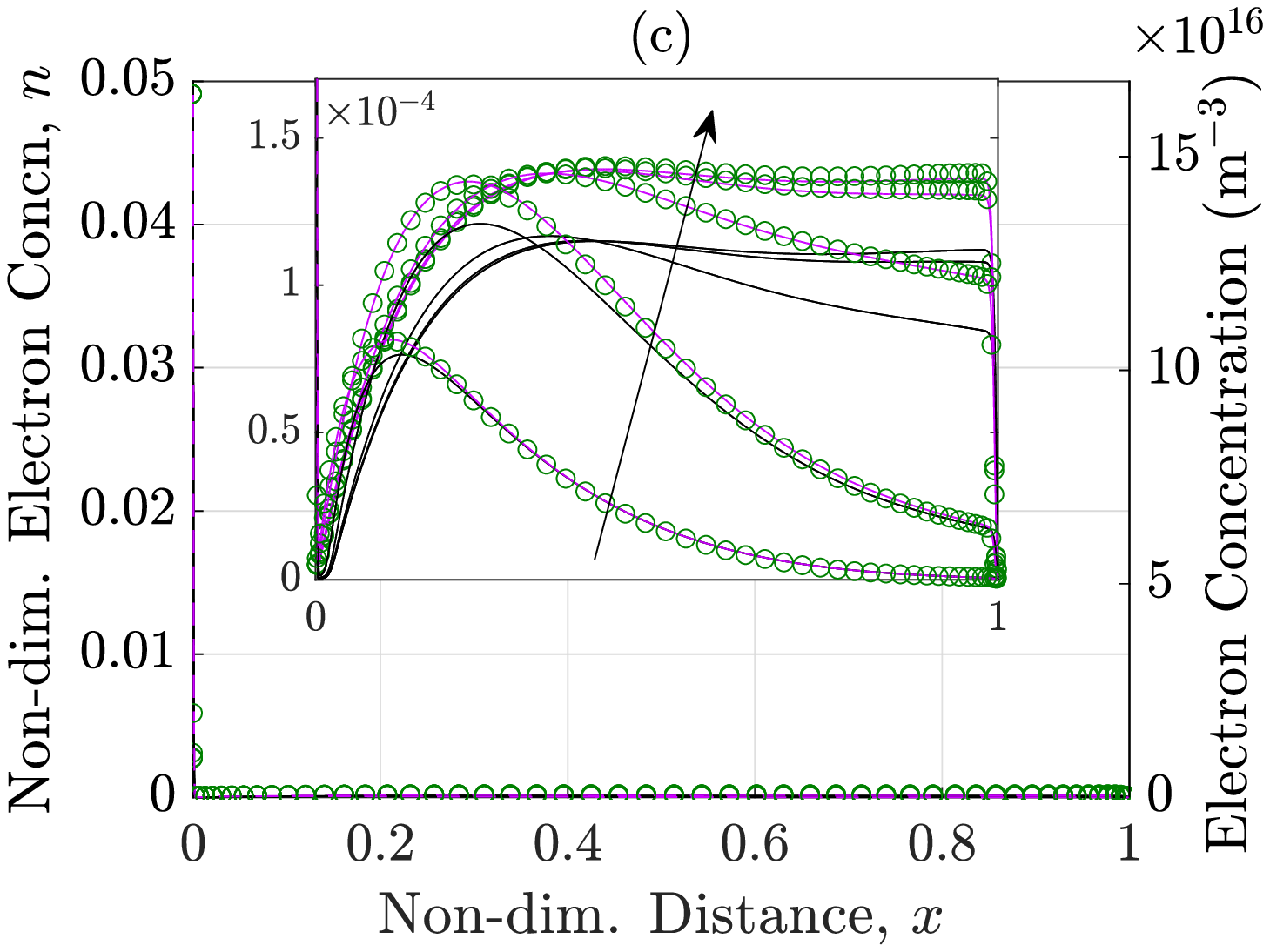}
\includegraphics[width=0.49\textwidth]{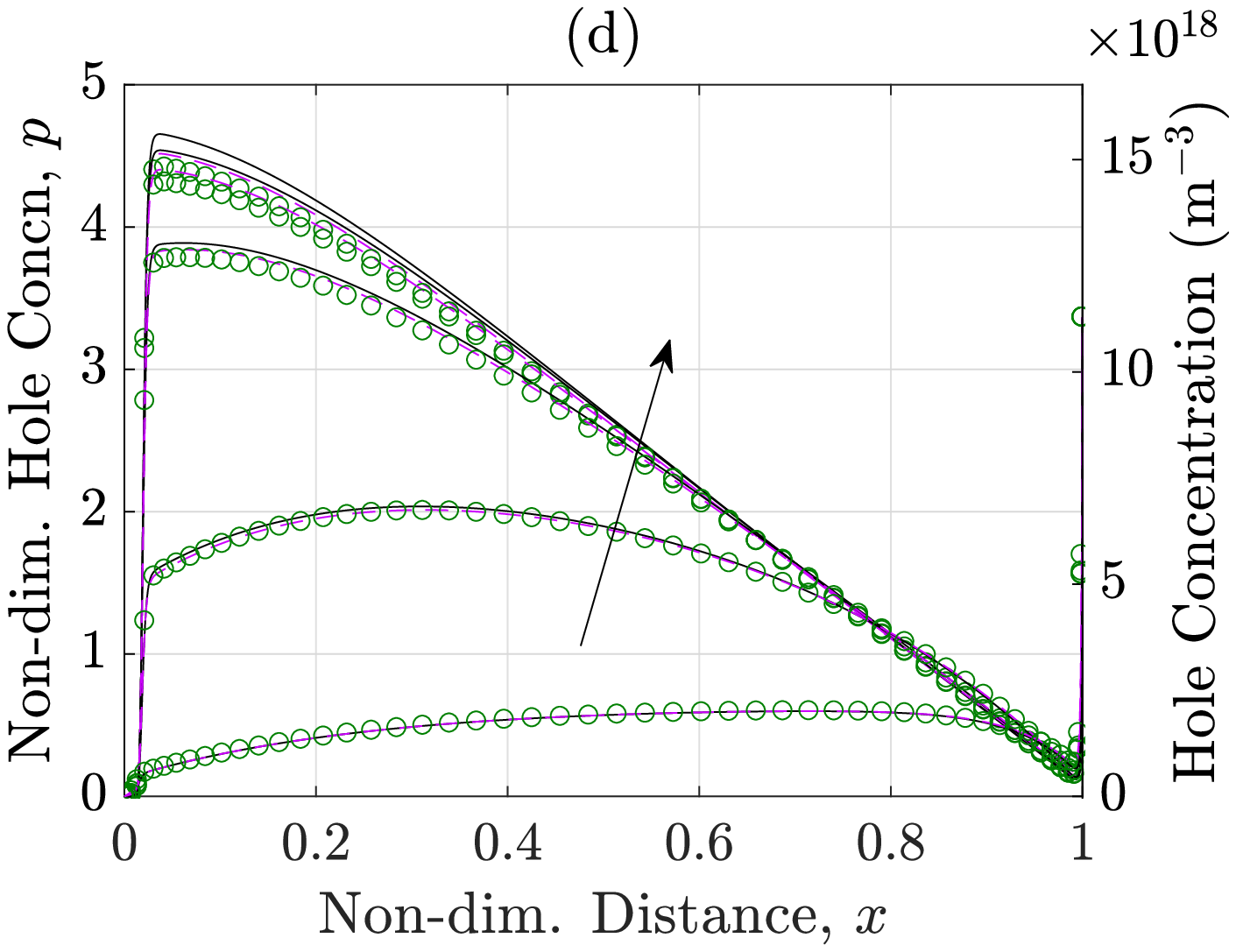} 
\caption{ As for figure \ref{FastEvol-08V} but for fast evolution in applied bias from $V_{ap}=V_{bi}$ to 0V.}
\label{FastEvol-0V}
\end{figure*}

\begin{figure*}
\includegraphics[width=0.49\textwidth]{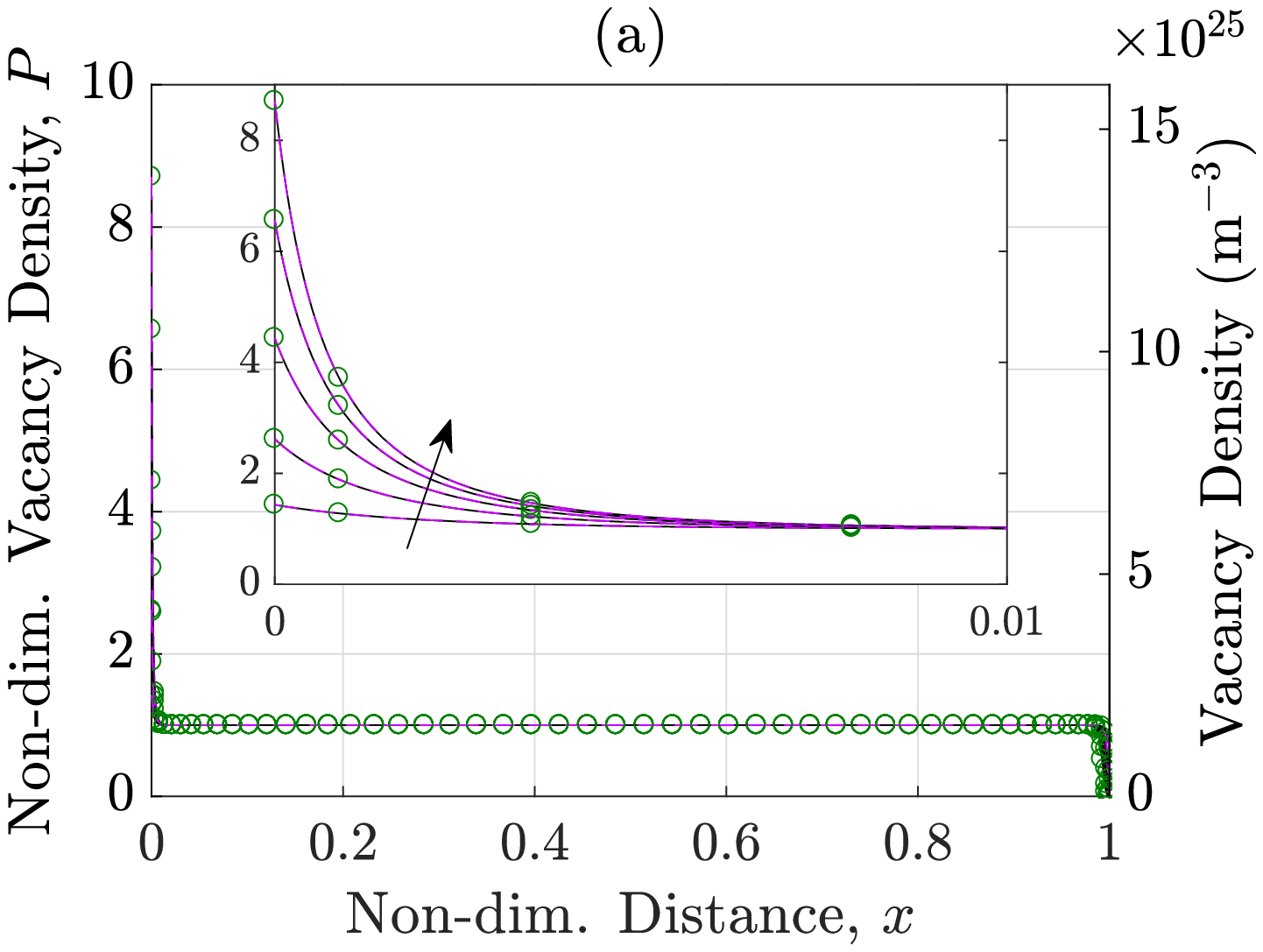}
\includegraphics[width=0.49\textwidth]{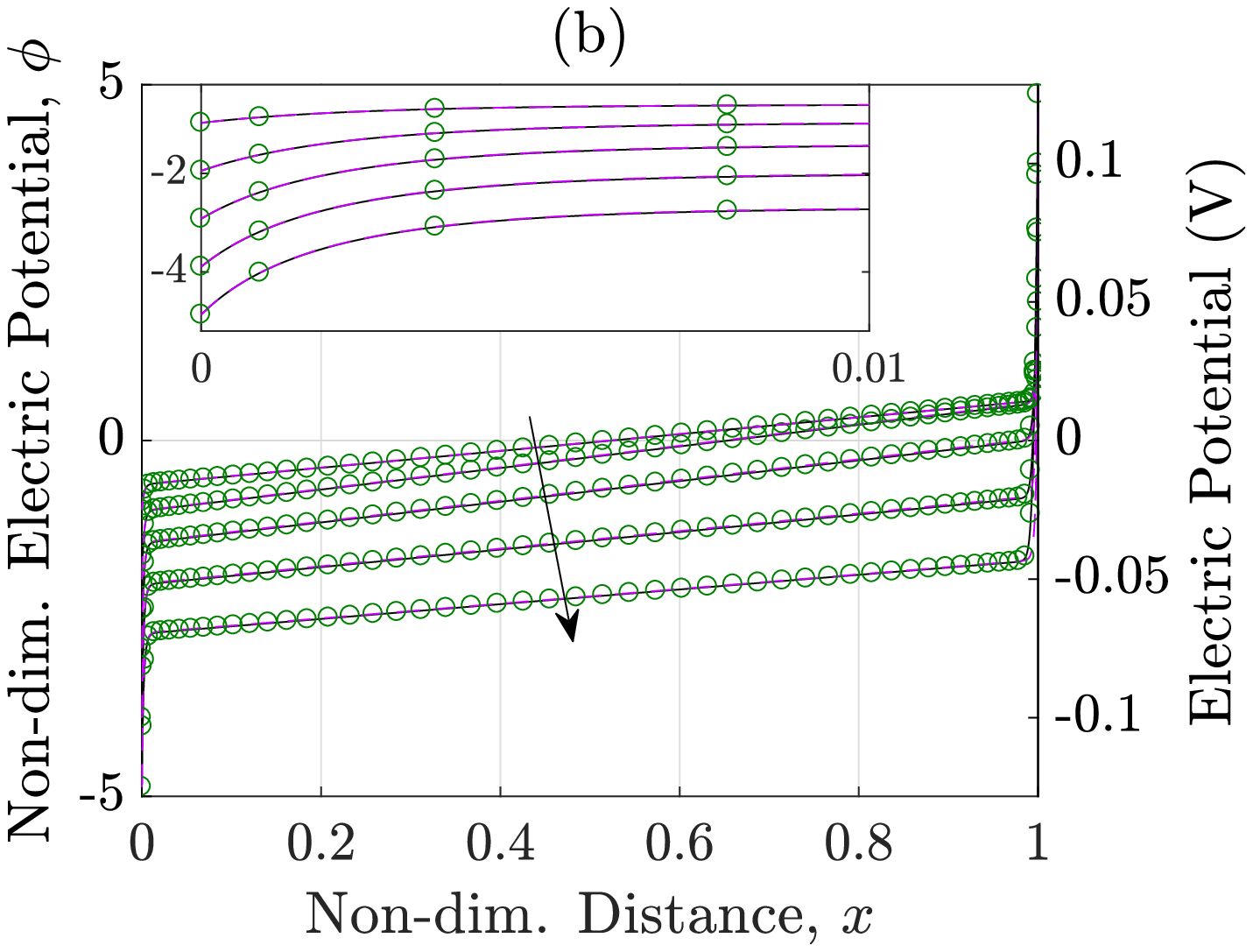} 
\includegraphics[width=0.49\textwidth]{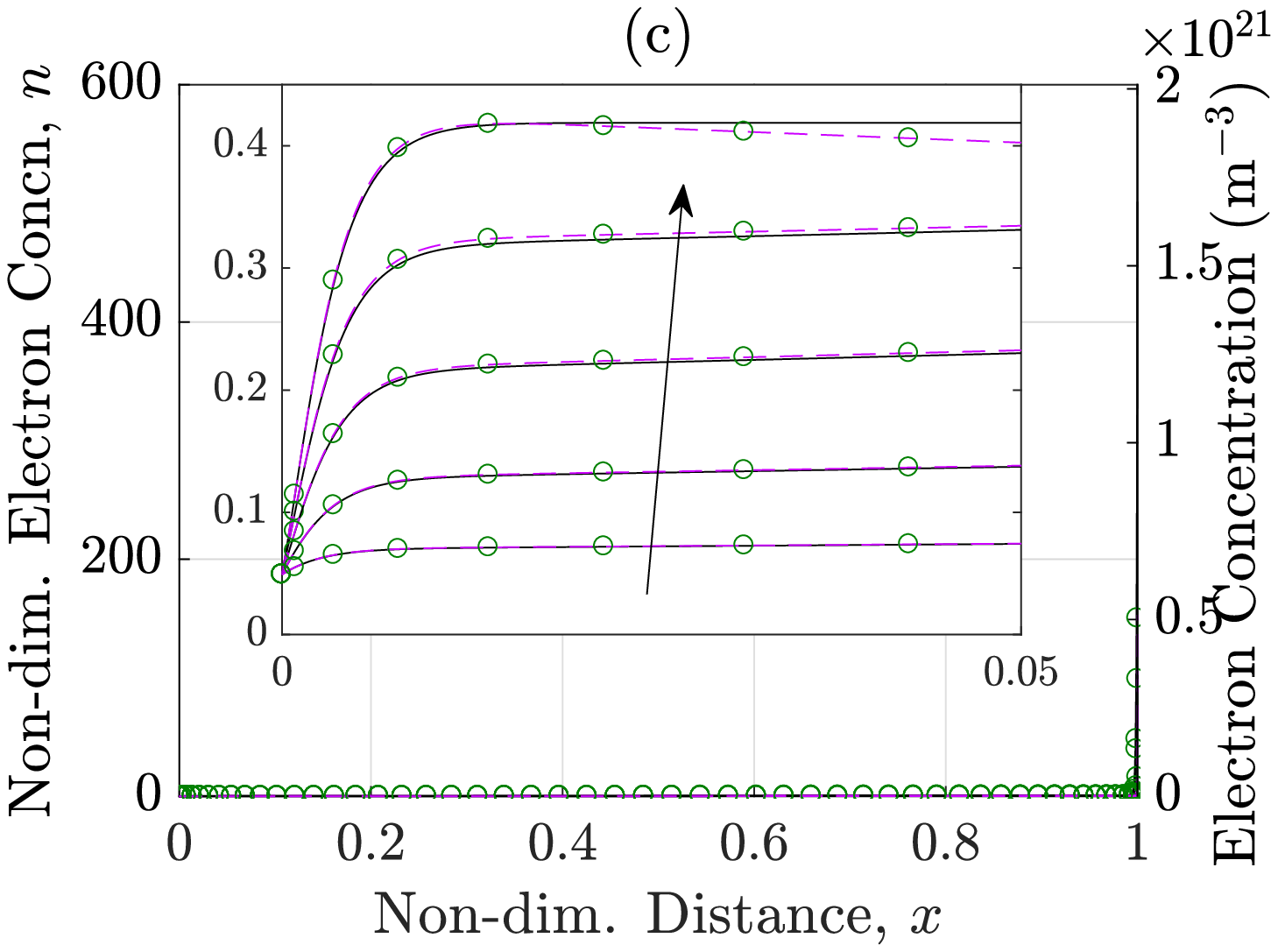}
\includegraphics[width=0.49\textwidth]{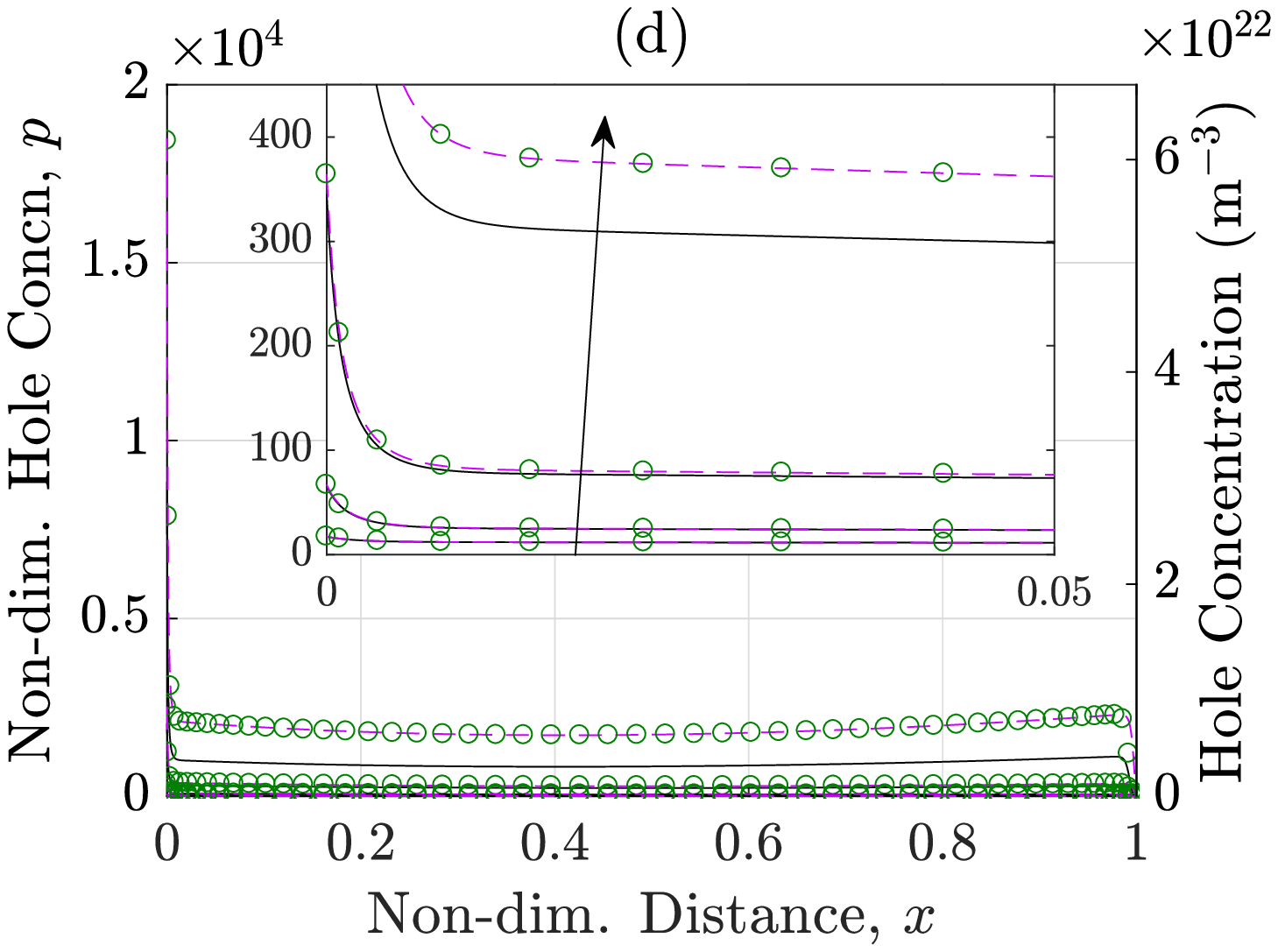} 
\caption{As for figure \ref{FastEvol-08V} but for a slow increase in applied bias from $V_{ap}=V_{bi}$ to 1.2 V.}
\label{SlowEvol-12V}
\end{figure*}

\begin{figure*}
\centering
\includegraphics[width=0.49\textwidth]{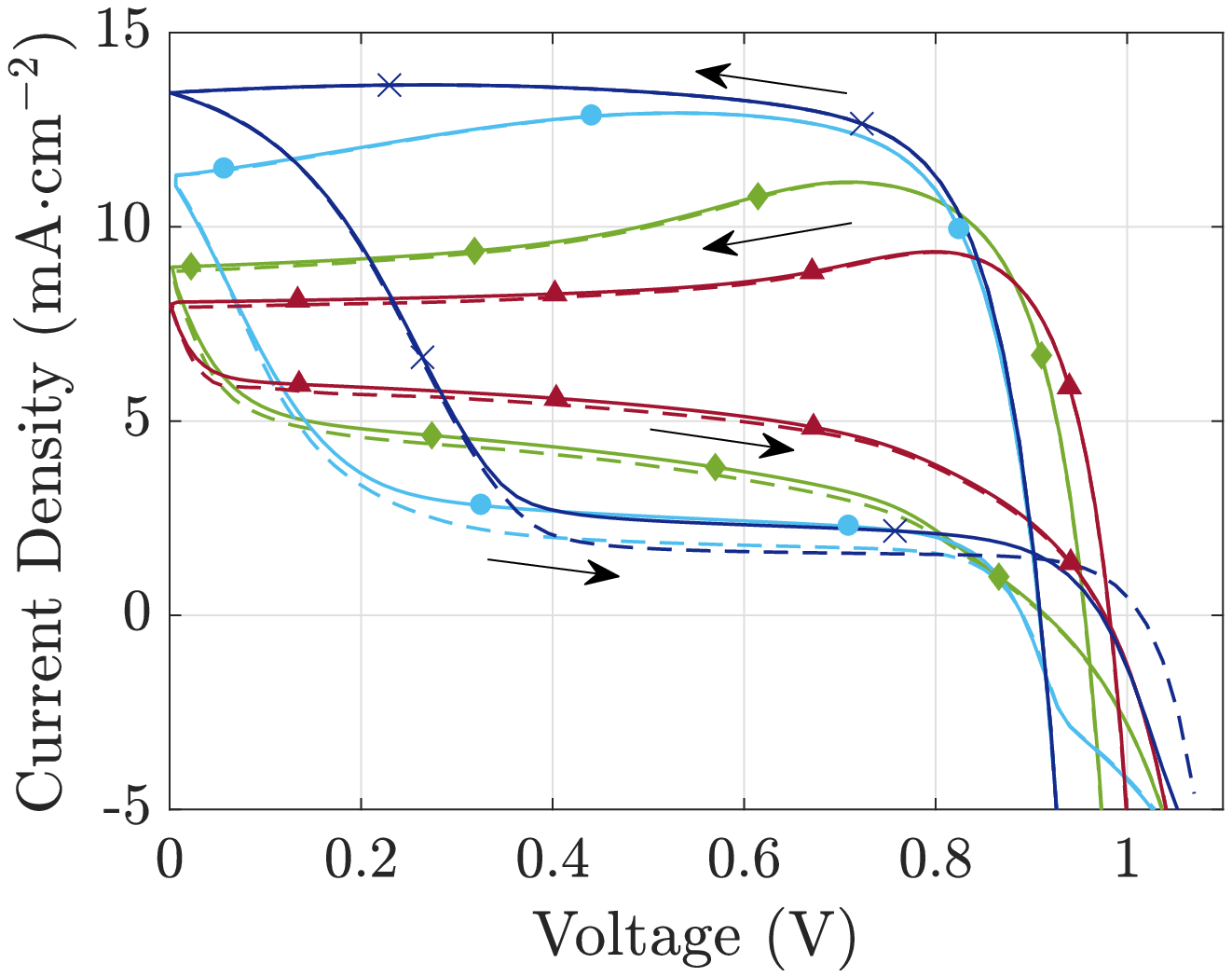}
\includegraphics[width=0.49\textwidth]{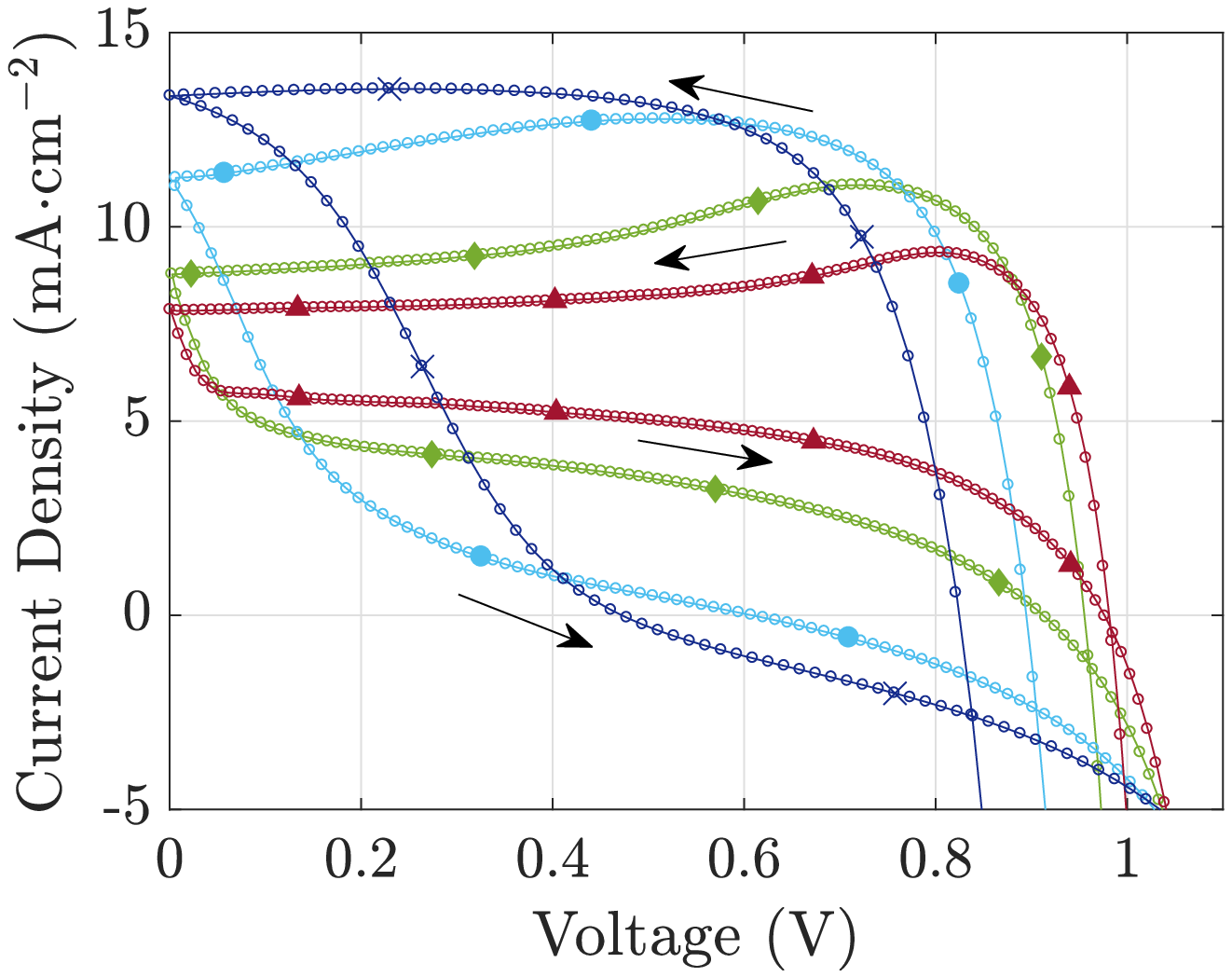}
\caption{Comparison of solutions for J-V curves at 4 different scan rates: 500mVs$^{-1}$, blue with crosses; 250mVs$^{-1}$, cyan with circles; 100mVs$^{-1}$, green with diamonds; and 50mVs$^{-1}$, red with triangles. Arrows show the direction of the voltage sweep. }
\label{JVCurves}
\end{figure*}

In figures \ref{FastEvol-08V}-\ref{JVCurves}, we show results for a device characterised by the parameters given in Table \ref{parameters table} with the perovskite layer width equal to 600 nm, corresponding to the set of dimensionless parameters given in (\ref{param-estim}). All numerical calculations are performed on a spatial grid consisting of 800 points.

Figures \ref{FastEvol-08V}-\ref{SlowEvol-12V} show the internal state of the cell at five equally-spaced values of time during a variation of the applied voltage, in a scenario in which the cell is abruptly illuminated at $t=0$s having been preconditioned in the dark with $V_{ap}=V_{bi}$. For figure \ref{FastEvol-08V}, the applied bias is varied smoothly from $V_{ap}=V_{bi}$ at $t$=0 s and $V_{ap}$ = 0.8 V at $t$=10 s (precisely, $V_{ap} = V_{bi}-0.2\tanh{(t)}/ \tanh{(10)}$).
For figure \ref{FastEvol-0V}, the applied bias is instantaneously decreased from $V_{ap}=V_{bi}$ to $V_{ap}=0$V at $t=0$s and held there for 4 seconds. Plots show solutions at $t=0.8,1.6,2.4,3.2,4.0$ s. Finally, for figure \ref{SlowEvol-12V}, the applied bias is varied linearly from $V_{ap}=V_{bi}$ at $t$=0 s to $V_{ap}$ =1.25V at $t$=10s. Plots are for $t=2,4,6,8,10$ s.

In figure \ref{JVCurves}, comparison is made between current-voltage (J-V) curves calculated using all three approaches and which model the experimental data presented by Richardson \etal\cite{Richardson16a}. Here, the cell is preconditioned for 5 seconds at 1.2V in the light before the J-V curve is measured. The current is calculated at equally-spaced intervals in time as the applied voltage is varied at a constant rate from 1.2V (forward bias) to 0V (short-circuit) and back; the four different scan rates are 50mVs$^{-1}$, 100mVs$^{-1}$, 250mVs$^{-1}$ and 500mVs$^{-1}$. The colour scheme has been chosen for consistency with Fig. 7 (b) from Richardson \etal\cite{Richardson16a}. In panel (a), solutions calculated using (i) the fully numerical (solid lines) and (ii) the combined asymptotic/numerical approach (dashed lines) are shown. Note that both of these methods calculate currents based on the full SRH recombination rate, (\ref{SRHequation}). While in panel (b), solutions from (iii) the fully asymptotic approach are shown.

\subsection{Discussion}

We have looked at three approaches to solving the drift-diffusion model (\ref{nondimDD})-(\ref{nondimics}). Approach (i) is fully numerical and involves solution of the full problem. In contrast, in approach (ii) (used previously\cite{Richardson16a}), we formally take the limits $\lambda \ra 0$ (small Debye length) and $\delta \ra 0$ (charge carrier concentration negligible in comparison to ion vacancy concentration). The comparison between the results of these two approaches is extremely favourable, as illustrated by the very small discrepancies in the J-V curves calculated using both approaches, for a range of scan rates, in figure \ref{JVCurves}(a).

The other main approximations to the drift-diffusion model that we make use of are the quasi-steady carrier limit, $\nu \to 0$ and the approximation of SRH recombination by hole dominated monomolecular recombination, $\epsilon \to 0$. The former limit ($\nu \to 0$) and its use, or otherwise, makes negligible difference to the results obtained. The latter, however, is frequently problematic, despite the very small value of $\eps$ ($=1/300$) we use in the simulations. This slightly surprising result is best illustrated by the significant differences between J-V curves calculated using the fully numerical method (solid curves in figure \ref{JVCurves}(a)) and those calculated using the fully asymptotic method in the limit $\eps \ra 0$ (figure \ref{JVCurves}(b)). Where there are significant differences between the two approaches this can be ascribed to strong spatial variations in charge carrier concentrations across the cell, resulting in regions where $n \leq O(\epsilon \, p)$ so that the approximation of $R(n,p)$ in (\ref{recombgennd}) by $R(n,p) \approx \gamma p$ no longer holds.


\section{Conclusion\label{concl}}
In this work we outlined a model for charge carrier transport and ion vacancy motion in a tri-layer planar perovskite solar cell (previously discussed in Richardson \etal \cite{Richardson16a}). Using parameters extracted from the literature, we were able to identify two key small dimensionless parameters that characterise the model: $\lambda$, which gives the ratio of the Debye length in the perovskite to the width of perovskite layer, and $\delta$, the ratio of the typical charge carrier (electron and hole) densities to the typical ion vacancy density. Based on the small size of these parameters, we performed an asymptotic analysis of the model which showed that: (a) the problem for the ion vacancy density and the electric potential is almost completely independent of the charge carrier densities and (b) the decoupled problem for ion vacancies and electric potential is well-approximated by the solution to a single first order ODE that describes the evolution of charge in the Debye layers (at the edge of the perovskite) in terms of the current through a resistor and a nonlinear capacitor in series. In dimensional form, this simplified model states that the charge (per unit area) in the right-hand Debye layer, $\Qp$, evolves according to the equation
\be
\frac{d \Qp}{d t}= \frac{q D_+ N_0}{V_T} \left(\frac{V_{bi}-V-\V(\Qp)+\V(-\Qp)}{b} \right)
\ee
where the term in the brackets is the (uniform) electric field in the perovskite bulk (away from the Debye layers) and $\V(Q)$ is the inverse to the nonlinear capacitance relation
\be
Q(\V)=\frac{\vep_p V_T}{L_d} \sign(\V) \left[ 2 \left( \exp(\V/V_T) -1-\frac{\V}{V_T} \right) \right]^{1/2} \;.
\ee
A good approximation to the full model can then be obtained by solving this much simplified problem for ion vacancy density and electric potential and using the resulting electric potential as an input into the charge carrier equations. The resulting model can sensibly be termed a surface polarization model of charge transport because it describes the effect on the current in a cell of the polarization of the perovskite layer, as ionic charge is transported from one of its surfaces to the other. In general, the simplified problem that we are left to solve for the charge carrier densities is nonlinear and so requires numerical solution. However, in contrast to the problem for the ion vacancies and potential, it is non-stiff and so this is not usually problematic. Moreover, parameter estimates suggest that the Shockley-Read-Hall recombination term in the charge carrier equations can be well approximated by monomolecular hole dominated recombination ($R(n,p) \approx \gamma p$). This allows the charge carrier equations to be linearised and, in turn, solved analytically. Where this is the case, an asymptotic solution to the entire model can be obtained from the solution to the single first order ODE discussed above.

In order to test the validity of the asymptotic method used to solve this model, we compared our asymptotic results to the results of a numerical solution of the full model. The latter was conducted using a recently developed numerical procedure\cite{Courtier17} that is able to accurately solve the full model in realistic parameter regimes. Where we used a combined asymptotic/numerical approach (solving for the ion vacancy distribution and electrical potential using the asymptotic model and solving for the charge carrier densities and currents numerically), we found extremely good agreement to the full numerical solution. In the case where we additionally linearised the charge carrier equations and solved them analytically, the comparison to the full numerical solution, while still good, was less impressive.

The physics of perovskite solar cells is still far from fully understood and in order to improve this situation it is vital that drift-diffusion models and their solution techniques continue to be developed. One obvious, and important, extension to the model discussed here is the explicit inclusion of charge transport in the electron- and hole-transport layers on either side of the perovskite. Such an extension will be able to elucidate how the choice of these layers affects the cell's transient behaviours. In particular, this extended model could be used to investigate cell architectures giving rise to {\it so-called} low hysteresis behaviour and would also be better able to account for interfacial recombination, see for example \cite{Calado16,Tan17}. Here we assume cation vacancies are immobile, which is justified by the relatively short timescales. However, it is believed that mobile methylammonium vacancies can lead to slow (over the timescale of many hours) but reversible changes in efficiency \cite{Domanski17}.

\section*{Author Contributions} NEC and JMF coded the numerical solver and produced the plots. GR performed the asymptotic analysis. GR and NEC wrote the manuscript. ABW, SEJO'K and GR conceived the project and formulated the model.

\section*{Acknowledgements}NEC is supported by an EPSRC funded studentship from the CDT in New and Sustainable Photovoltaics. SEJO'K was supported by EPSRC grant EP/J017361/1. ABW acknowledges funding from the European Union Horizon 2020 research and innovation programme under grant no. 676629.

\bibliographystyle{siam}
\bibliography{courtierprb17bib}

\appendix
\section{\label{append}Solution for $P$ and $\phi$ in the Debye layers}
In the bulk region, we obtain a solution for the leading order vacancy density $\poz$ and potential $\phoz$, given by $\poz=1$ and $\phoz=W_-(t)(1-x)+W_+(t) x$. These expressions satisfy the potential boundary conditions but in general cannot satisfy the flux boundary conditions, see (\ref{ionDDbc}). In order to resolve this seeming paradox, we need to account for narrow boundary layers (Debye layers) of width O($\lambda$) about $x=0$ and $x=1$.
\paragraph*{Debye layer about $x=0$.} Considering first the Debye layer about $x=0$, we use the rescaling (\ref{zeta-def}) to rewrite the governing equations (\ref{ionDDeq})-(\ref{ionDDbc}) in terms of the rescaled spatial variable $\zeta$, yielding the boundary layer equations:
\be
\frac{\dd P}{\dd t} &+& \frac{\dd \fp}{\dd  \z} = 0\;, \quad
\fp=- \frac{1}{\lambda}\left(\frac{\dd P}{\dd \z} +P \frac{\dd \phi}{\dd \z}   \right), \label{bleqs} \\
\frac{\dd^2 \phi}{\dd \z^2}&=&  \left( 1-P \right), \label{blPoi} \\
\phi|_{\z=0} &=&\frac{\Phi-\Phi_{bi}}{2}, \quad \left. \fp \right|_{\z=0}=0, \quad P|_{t=0}=1. \label{blbcs}
\ee
The expansions for $P$, $\phi$ and $\fp$ proceed as in (\ref{deb-exp1}) so that to leading order in (\ref{bleqs}) we obtain the following equation for $\pdz$.
\bes
\frac{\dd \pdz}{\dd \z} +\pdz \frac{\dd \phidz}{\dd \z}=0,
\ees
This has the solution
\be
\pdz=\exp(W(t)-\phidz), \label{p-debye}
\ee
for some as yet undetermined function of time, $W(t)$.

\paragraph*{Matching to the outer.} In order for the leading order Debye layer solution to match to the leading order outer solution, through (\ref{ionvar})-(\ref{ionconcexp}), we require
\be
\pdz \ra 1, \quad \phidz \ra W_-(t), \quad  \z \ra +\infty. \label{match1}
\ee

Applying the matching condition (\ref{match1}) to the solution (\ref{p-debye}) determines a relation between the arbitrary functions $W(t)=W_-(t)$ motivating us to eliminate one of them by writing
\be
\pdz=\exp(W_-(t)-\phidz). \label{p-debye2}
\ee
On substituting this expression into (\ref{blPoi}) balanced at leading order, we find
\be
\frac{\dd^2 \phidz}{\dd \z^2}= \left( 1-  \exp(W_-(t)-\phidz)  \right), \label{ddphi}
\ee
which satisfies boundary conditions obtained from the leading order terms in (\ref{blbcs}) and from (\ref{match1}), namely
\be
 \phidz|_{\z=0}= \frac{\Phi_{bi}-\Phi}{2}, \quad \phidz \ra W_-(t),\; \z \ra +\infty. \label{phibc}
\ee
The corresponding expansion for the total charge per unit area in the left-hand Debye layer, $\Qm$ (defined in (\ref{qm-def})) is
\bes
\Qm=\Qmz+\cdots,
\ees
and, by substituting this into (\ref{qm-def}), we obtain
\be
\Qmz= \int_0^{\infty} \left(\exp(W(t)-\phidz) - 1\right) {\rm d} \z. \label{qmz}
\ee

We can reformulate the problem for $\phidz$, given by (\ref{ddphi})-(\ref{phibc}), in a generic form by writing 
\be
\phidz(\z,t)=\theta(\z,\Vm(t))+W_-(t), \label{theta-def}
\ee
where $\Vm(t)$ represents the potential gained across the Debye layer, \ie $\Vm=[\phidz]_0^{\infty}$ (note that with this definition $\theta=-\log_e \pdz$). It is then straightforward to show that the function $\theta(z,\V)$ must satisfy the generic modified Poisson-Boltzmann problem
\be
&&\frac{\dd^2 \theta}{\dd z^2}= \left( 1-  e^{-\theta}  \right), \label{thetaeq}\\
&&\theta|_{z=0}=-\Vm(t), \quad \theta \ra 0 \text{ as }  z \ra +\infty. \label{thetabc}
\ee
Furthermore, in order that $\phidz|_{\z=0}=\frac{1}{2} ( \Phi_{bi}-\Phi)$,
\be
W_-(t)=\frac{\Phi_{bi}-\Phi}{2}+\Vm(t).
\ee
Thus if we are able to determine $\Vm(t)$, we can determine the unknown function $W_-(t)$ in the leading order outer solution for the potential in (\ref{ionconcexp}). 

It is straightforward to obtain a first integral to the autonomous equation (\ref{thetaeq}) in the standard fashion by multiplying by $\theta_{z}$ and integrating with respect to $z$. This yields, on applying the far-field condition (\ref{thetabc}), the expression
\be
\frac{\dd \theta}{\dd z}= \sign(\V) \sqrt{2} \left(\theta+e^{-\theta}-1 \right)^{1/2}, \label{first-intg}
\ee
where the $-\sign(\V)$ is to account for the fact that if $\V<0$ ($\V>0$) the gradient of $\theta$ must be negative (positive). We can integrate (\ref{first-intg}) once more to obtain a relation for $z$ as a function of $\theta$ which reads
\be
\begin{array}{ccccc}
z=\Zo \left(\theta \right)- \Zo(-\V) & \mbox{for} & \V>0 & \mbox{with} & \theta<0, ~~\\
z=\Zt \left(\theta \right) - \Zt(-\V) & \mbox{for} & \V<0 & \mbox{with} & \theta>0, ~~
\end{array} \label{theta-reln}
\ee
where
\be
\begin{array}{ccc}
 \Zo(\theta)= \frac{1}{\sqrt{2}} \int_{-1}^{\theta} \frac{ {\rm d} w}{(w+e^{-w}-1)^{1/2}}  & \mbox{with} & \theta<0,\\*[4mm]
 \Zt(\theta)= \frac{1}{\sqrt{2}} \int_{\theta}^{1} \frac{ {\rm d} w}{(w+e^{-w}-1)^{1/2}}  & \mbox{with} & \theta>0.
\end{array} \label{zozt}
\ee

\paragraph*{Characterising the capacitance of the Debye layer.}
We now seek to relate $\Qmz$, as given in (\ref{qmz}), to $\Vm$. We note that 
\bes
\Qmz=\int_{0}^{\infty} \left(e^{-\theta}-1\right) {\rm d} \z,
\ees
which we can rewrite as
\bes
\Qmz=\int_{-\Vm}^0  \frac{e^{-\theta}-1}{\theta_{\z}}  {\rm d} \theta.
\ees
On substituting for $\theta_{\z}$ from (\ref{first-intg}) and writing $\theta=-\V$ this integral transforms to
\bes
\Qmz=\frac{\sign(\Vm)}{\sqrt{2}} \int_0^{\Vm}  \frac{e^{\V}-1}{(e^{\V}-\V -1)^{1/2}}  {\rm d} \V.
\ees
This integral can be further transformed by the substitution $M(\V)=e^{\V}-\V -1$ to the exact integral
\bes
\Qmz=\frac{\sign(\Vm)}{\sqrt{2}} \int_0^{M(\Vm)} \frac{1}{M^{1/2}}  {\rm d} M
\ees
which yields the following exact relation between $\Qmz$ and $\Vm$:
\be
\Qmz = \sign(\Vm) \left( 2(e^{\Vm}-\Vm -1) \right)^{1/2}.  \label{capac1}
\ee
This relation is plotted in figure \ref{capac_plot}, from which it can be seen that $\Qmz$ is a single valued function of $\Vm$. Hence, given the Debye layer charge density, $\Qmz$, we can invert to find the potential jump across the Debye layer, $\Vm$. This motivates us to consider the evolution of $\Qmz(t)$ as charge (in the form of positively charged vacancies) flow out into (or in from) the bulk region.

\paragraph*{A solvability condition on $\Qmz(t)$.} It remains to determine the evolution of $\Vm(t)$. This can be done by tracking the charge build up in the Debye layer through the leading order expansion of the positively charged vacancy conservation equation (\ref{bleqs}),
\be
\frac{\dd \pdz}{\dd t}+ \frac{\dd \fpdz}{\dd  \z}=0, \label{dpdzdt}
\ee
and the boundary conditions 
\be
\fpdz|_{\z=0}=0, \quad \fpdz \ra \fpoz|_{x=0} \text{ as } \z \ra +\infty. ~~\label{pdzbc}
\ee
These conditions are obtained from the leading order expansion of (\ref{blbcs}b) and from matching to the leading order outer solution as $\z \ra+ \infty$, respectively. By writing 
${\dd \pdz}/{\dd t}$ as $({\dd }/{\dd t}) (1-\pdz)$, integrating (\ref{dpdzdt}) between $\z=0$ and $\z=\infty$ and applying the flux boundary conditions (\ref{pdzbc}), we obtain the solvability condition
\be
\frac{d \Qmz}{d t} = -\fpoz|_{x=0}.
\ee

\paragraph*{The Debye layer about $x=1$.}
The analysis of this right-hand layer proceeds in a similar fashion to the left-hand Debye layer presented above. We introduce the rescaled spatial variable $\xi$, defined in (\ref{xi-def}), and then expand as follows.
\bes
P &=& \prz(\xi,t)+\cdots\;, \; \fp=\fprz(\xi,t)+\cdots, \\
\phi &=& \phirz(\xi,t)+\cdots\;, \; \Qp=\Qpz(t)+\cdots.
\ees
Following an analogous series of steps to the analysis of the left-hand layer, we find that
\be
\prz=\exp\left(W_+(t)-\phirz\right) \;,
\ee
and that the leading order potential satisfies the problem
\be
&&\frac{\dd^2 \phirz}{\dd \z^2} = 1-  \exp\left(W_+(t)-\phirz\right)  \;, \nonumber \\
&&\phirz|_{\xi=0} = -\frac{\Phi_{bi}-\Phi}{2} \;, \nonumber \\
&&\phirz \ra W_+(t) \text{ as }  \xi \ra \infty.
\ee
The solution to this problem is very similar to that for $\phidz(\z,t)$ being given by
\be
\phirz(\xi,t)=\theta(\xi, \Vp(t))+W_+(t) \;,  \label{thetab-def}
\ee
where $\Vp(t)=[\phirz]_{\xi=0}^{\infty}$ is the jump in potential across the right-hand Debye layer and the function $\theta(\cdot,\cdot)$ is (as before) a solution to (\ref{thetaeq})-(\ref{thetabc}); in other words, one can make the transformation $\eta\ra \xi$ and $\Vm \ra \Vp$ in $\theta(\xi,\Vm)$ both here and in the implicit solution for $\theta$ given in (\ref{theta-reln}). In addition, it follows from the condition that $\phirz|_{\xi=0}=-\frac{1}{2} (\Phi_{bi}-\Phi)$ that
\be
W_+(t)=-\frac{\Phi_{bi}-\Phi}{2} + \Vp(t).
\ee 

In a similar manner to that described above, we determine a relation between $\Qpz$ and $\Vp$ which is identical to (\ref{capac1}) and reads
\be
\Qpz = \sign(\Vp) \left( 2(e^{\Vp}-\Vp -1) \right)^{1/2}. \label{capac2}
\ee

Once again, a solvability condition may be derived from the problem for the leading order anion vacancy density, namely
\be
\frac{\dd \prz}{\dd t}- \frac{\dd \fprz}{\dd  \xi} = 0 \;,
\ee
\be
\fprz|_{\xi=0} = 0, \quad\fprz \ra \fpoz|_{x=1} \text{ as } \xi \ra +\infty. ~~~
\ee
The solvability condition we obtain on integrating this system is the following evolution equation for $\Qpz(t)$:
\be
\frac{d \Qpz}{d t}= \fpoz|_{x=1}.
\ee

\end{document}